\documentclass{article}

\usepackage{arxiv}
\usepackage{subcaption}
\usepackage{multirow}
\usepackage[utf8]{inputenc} 
\usepackage[T1]{fontenc}    
\usepackage[hidelinks]{hyperref}       
\usepackage{url}            
\usepackage{booktabs}       
\usepackage{amsfonts}       
\usepackage{nicefrac}       
\usepackage{microtype}      
\usepackage{lipsum}		
\usepackage{graphicx}
\usepackage{authblk}
\usepackage{comment}
\usepackage[sorting=none]{biblatex} 
\usepackage{doi}

\usepackage{newfloat}
\DeclareFloatingEnvironment[name={Supplementary Figure},fileext=lsf,listname={List of Supplementary Figures}]{suppfigure}

\providecommand{\keywords}[1]
{
  \small	
  \textbf{\textit{Keywords---}} #1
}
\title{On the Visualisation of Argumentation Graphs to Support Text Interpretation }

\date{} 					

\author[ ,1]{Hanadi Mardah\thanks{Corresponding author: hanadi.mardah@manchester.ac.uk. \\ PhD. student and a lecturer at Um Al Qura University email \emph{homerdah@uqu.edu.sa}. websites: https://uqu.edu.sa/en/Profile/homerdah or https://hanadi-mardah.sitelio.me/}} 
\author[1,2]{Oskar Wysocki}
\author[1] {Markel Vigo}
\author[1,2,3]{Andr\'{e} Freitas}

\affil[1]{Department of Computer Science, The University of Manchester, Manchester, United Kingdom}
\affil[2]{digital Experimental Cancer Medicine Team, Cancer Biomarker Centre, CRUK Manchester Institute, United Kingdom}
\affil[3]{Idiap Research Institute, Martigny, Switzerland}

\hypersetup{
pdftitle={A template for the arxiv style},
pdfsubject={q-bio.NC, q-bio.QM},
pdfauthor={David S.~Hippocampus, Elias D.~Striatum},
pdfkeywords={First keyword, Second keyword, More},
}

\begin{document}
\maketitle

\begin{abstract}
	The recent evolution in Natural Language Processing (NLP) methods, in particular in the field of argumentation mining, has the potential to transform the way we interact with text, supporting the interpretation and analysis of complex discourse and debates. Can a graphic visualisation of complex argumentation enable a more critical interpretation of the arguments? This study focuses on analysing the impact of argumentation graphs (AGs) compared with regular texts for supporting argument interpretation. We found that AGs outperformed the extrinsic metrics throughout most UEQ scales as well as the NASA-TLX workload in all the terms but not in temporal or physical demand. The AG model was liked by a more significant number of participants, despite the fact that both the text-based and AG models yielded comparable outcomes in the critical interpretation in terms of working memory and altering participants' decisions. The interpretation process involves reference to argumentation schemes (linked to critical questions (CQs)) in AGs. Interestingly, we found that the participants chose more CQs (using argument schemes in AGs) when they were less familiar with the argument topics, making AG schemes on some scales (relatively) supportive of the interpretation process. Therefore, AGs were considered to deliver a more critical approach to argument interpretation, especially with unfamiliar topics. Based on the 25 participants conducted in this study, it appears that AG has demonstrated an overall positive effect on the argument interpretation process.
\end{abstract}

\begin{figure}[h!]
\centering
  \fbox{\includegraphics[width=\textwidth]{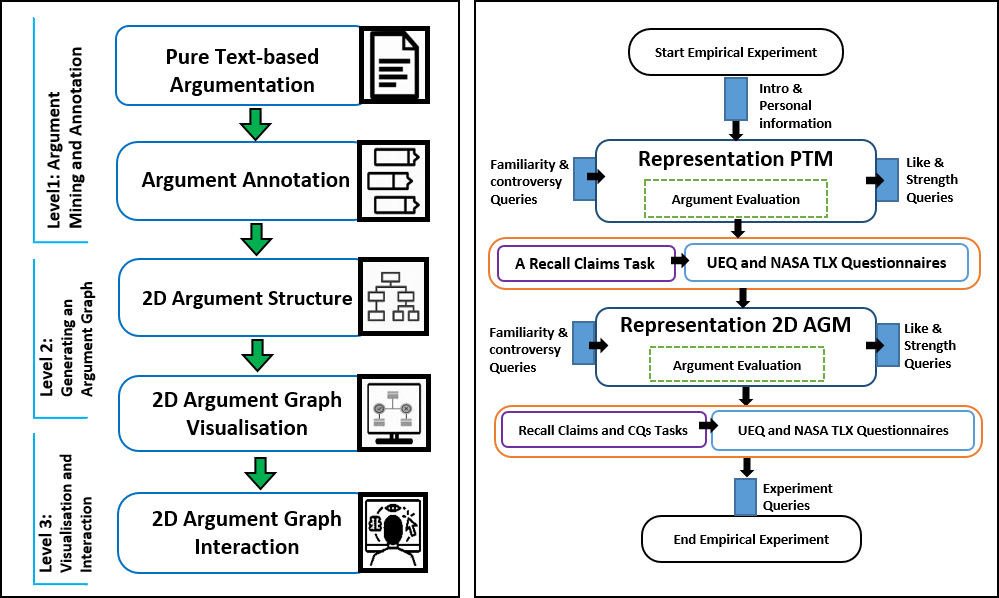} }
  \caption{(left) The study process, based on DAGGER framework \cite{Yun2018} and (right) the abstract empirical experiment workflow, more details Fig. \ref{fig:workflow}}
  
  \label{fig:experiment_workflow}
\end{figure}

\section{Introduction}
 Argumentative texts are a very prevalent type of discourse, and they consist of the process of persuading others to accept a certain attitude or opinion. The term ``argument'' refers to the subset of sentences that includes two main types of discourse units (sentences): ``premises'' and ``claims'' (in some cases called  ``conclusion''). These sentences are linked to each other in feasible chains of reasons offered in support of or to attack a claim, or for deriving further claims \cite{Walton2016}.  An argument's final claim is usually called its conclusion, a proposition that can be either true or false, put forward by someone as true, see a Supp. Fig. \ref{fig:Arg-str}.
 
 The argumentation structure is represented as a defeasible premises-conclusion structure and a set of CQs, which help in the procedure of testing the strength and acceptability of the argumentation, based on the weighting of the pro and con arguments. Automatically identifying an argument and its relevant components (claims and premises) in a text is called argument mining or argument extracting. This is a sub-area within natural language processing (NLP) that is rapidly evolving and has the support of multiple annotated corpora \cite{Tamames2010, Stede13}. With the support of these classifiers, arguments can be parsed to build a supporting graph representation, which is called an argument map, a visual representation of the logical argumentation structure, see Supp. example \ref{fig:Arg-ex}. An argument map or argument graph enables a complex argument to be broken down into isolated argument units, which can potentially facilitate the interpretation and analysis of arguments \cite{CoInfRes}.
 
This study investigates the impact of visual representations of argumentation graphs (AGs) on the interpretation of arguments. In particular, it focuses on the impact of visualisation and, to a minor extent, of interaction strategies on AG structures. Furthermore, we aim to reduce the barriers that hinder users’ understanding, analysis and debating of various existing arguments in a critical manner by proposing an argumentation visualisation model, developed by using a combination of the recent capabilities produced by argumentation mining. The study process framework (Fig. \ref{fig:experiment_workflow} (left)) is based on the framework of the DAGGER\footnote{DAGGER is a tool for generating argumentation frameworks from Datalog+/- inconsistent knowledge bases.} \cite{Yun2018} and includes three levels : (i) annotating using argument mining manually; (ii) generating AGs; and (iii) visualising and interacting with AGs. The study provides a critical evaluation of: (1) whether structured AG representations facilitate the interpretation of arguments compared to their textual counterparts; and (2) the impact of proposed AG on an argument interpretation task. Therefore, the study includes an empirical analysis, which compares a pure text-based model (\textbf{PTM}) as a baseline model with the proposed model, named 2D argumentation graph model (\textbf{2D AGM}). See Fig. \ref{fig:experiment_workflow} (right) and more details in Fig. \ref{fig:workflow}.

Pragmatically, this study aims to answer the following research questions (RQs), which are divided into two areas:
 \begin{enumerate}
     \item \textbf{Argumentation Representation:}
    \begin{itemize}
        \item RQ1: How do discourse-level graph representations support the interpretation of complex arguments and debates?\label{RQ1} 
        \item RQ2: Which categories and relationships positively affect the interpretation process?\label{RQ2}
        \item RQ3: How do argumentation schemes and argumentation types support the construction of this representation? \label{RQ3}
    \end{itemize}
 \item \textbf{Visualisation and Interaction:}
    \begin{itemize}
        \item RQ4: How does the proposed model perform for intrinsic and extrinsic evaluation metrics when compared to a pure text-based interpretation? \label{RQ4}
        \item RQ5: What are the users' perceptions of their preference for, and the easiness, effectiveness, and usefulness of the proposed solution? \label{RQ5}
        \item RQ6: How could the familiarity, strength, and controversy of topics affect the argument critical interpretation process? \label{RQ6}
    \end{itemize}
\end{enumerate} 

In summary, the contributions of this study are the following:

\begin{itemize}
    \item We proposed a 2D AGM using a node-link diagram. This is different from and complements the existing tools in visualisation and spatial configuration \cite{Brenneis2020, Green2019, cullen18, Rieche18, Baroni2015, Al-Shehhi2015}.
    
    \item As part of the methodology, we follow Walton’s classification schemes to represent the AG, which comprises three primary nodes (with different levels and colours) and four central relationships (with distinct colours). Then, we allow users to decide on the general argument topics based on the weights of the pros(+) and cons(-) of the central claims, using a visual scale tool. This is a contribution over other tools \cite{Baroni2015, Brenneis2020} in terms of how the explicit function of final decision mapping can support an intrinsic evaluation of argument components. Our hypothesis is that the 2D AGM will perform better in terms of the critical agreement evaluation of the argument (decision-making).
    
     \item We compare and contrast the main features of the critical argument interpretation process, such as time and working memory, between the two models. The experiment embeds tasks involving recalling claims and selecting critical questions (CQs). Using an A-BLEU score, we evaluate how similar the recalled claims are to the references in each model. Our hypothesis is that the 2D AGM will yield better outcomes in time and working memory compared to the PTM.
    \end{itemize}
    
    To achieve the objective of this study, we have to include the following:
    \begin{itemize}
    \item We conducted empirical analysis on 25 users who interacted with two model representations: PTM and 2D AGM. Each model included an agreement evaluation, tasks, and two questionnaires (NASA- TLX and UEQ). The experiment included a survey about the topics’ familiarity, controversy, and strength, and whether or not the user liked the model. It also included queries about models and the experiment. 
    
    \item We performed an extrinsic evaluation in terms of workload and user perception via two questionnaires and several queries about the familiarity, controversy, and strength of the argument topics. The queries determine the impact of these characteristics on the interpretation process across the two models. This helps to support a rational evaluation and provides a more intuitive interpretation for users. Our hypothesis is that the 2D AGM will achieve a superior workload and user perception compared to the PTM because the fundamental reason for using AGs is to break down a complex argument and facilitate interpretation. AGs will have a greater influence on the argument’s interpretation, depending on the familiarity of the topic, its controversy and its strength, compared to the PTM.
\end{itemize}
 
\section{Background and Literature Review}
Argumentation aims to justify conclusions and actions through rational and evidence-based beliefs, presenting disagreements, demonstrating truth, and understanding multiple perspectives \cite{Bembenik2016}. Despite advances in automatic argument structuring and visualizations in Natural Language Processing NLP and Argumentation Mining AM, there is a significant gap in understanding and quantifying to which extent this added structure, combined with visualisation methods, can support the human interpretation of complex arguments. 
For this reason, this study describes visualisation and interaction with argumentation graphs (VIAGs), which is a sub-theme of Computer-Supported Argumentation Visualisation (CSAV) \cite{Riechert18}. This section reviews and summarises approaches reported in the literature related to the representation of arguments and argument mapping/visualisation models (Fig. \ref{fig:experiment_workflow} (left)).

\subsection{Argumentation Structure and Walton's Classification Schemes}
 There is a challenge in categorizing and structuring arguments \cite{Lippi2016, Walton2016}). There is no universal definition of a `structured argument', making it a fundamental challenge associated with finding a classification system to identify different argument patterns and produce an argument for various circumstances and purposes\cite{Macagno2017}.  Several attempts have been made to classify schematic structures \cite{Hastings63Reformulation, Kienpointner1994, Walton1996, Grennan1997, Lawrence2016}, but the most commonly used scheme is given by Walton \cite{Walton2008}. According to a survey by Lawrence et al. \cite{Lawrence2019}, arguments can be evaluated based on critical questions corresponding to Walton's scheme. Can et al.\cite{Can2019} also found that Walton's model is useful in supporting argumentation and critical evaluation. Our study will use Walton's classification scheme \cite{Walton2008, Walton2016, Milz2017} to represent argumentative text.
\subsection{Argument Mining and Annotation Methods}
Argument mining (AM) is a technique in text mining \cite{Talib2016} that involves two main stages: identifying argument discourse and predicting argument relationships \cite{Lippi2016, Cabrio2018}. AM can be done fully automated (\cite{Trautmann, Swanson, Ma20, Stab, LippiMarco2016, Lawrence2016, Douglas2011, Noor, Hunter2020}, TARGER\cite{Chernodub}, MARGOT ARGs\cite{Wachsmuth2017b}, or ConToVi \cite{El-Assady2016}) or as a hybrid of manual and automated methods (\cite{Lenz2020, Lawrence2015}). However, it lacks standardization \cite{Lippi2015} and full automation lacks accuracy due to its insufficient incorporation of semantics and domain knowledge. Consequently, experts currently rely on time-consuming manual annotations \cite{Sperrle}. This study aims to refine Walton's classification schemes \cite{Walton2016} to analyze debate topics from ``ProCon.org'' \cite{procon} using the BRAT annotation tool \cite{brat} and the VIANA framework, with a focus on enhancing textual interpretation in argument mining.

\subsection{Argument Mapping or Visualizing}
Argument structure visualisation, also known as argument representation, mapping or graphs, is a form of text visualization \cite{Kucher2015}. Argument graphs can be displayed as node-link diagrams or trees, with node-link diagrams suitable for hierarchical structures\cite{Katifori2007, Walton2008}. According to \cite{Bembenik2016}, boxes/nodes usually contain full, grammatical, declarative sentences: reasons/premises (pieces of evidence in support of some claims); claims (ideas claimed to be true); conclusions (final claims supported by reasons); or objections (pieces of evidence against conclusions).  Relationships between nodes are shown with lines/arrows representing reasoning relationships. Therefore, this study constructs and represents Walton's argument structure as a tree (a node-link diagram) with three primary nodes: premises, claims, and conclusions, and four central relationships: attack, support, rebuttal, and undercutting\cite{Walton2016}, visualised as a 2D graph-based argument model, rendered using the Unity framework \footnote{Unity \cite{unity} is a cross-platform game engine developed by Unity Technologies, first announced and released in June 2005 at Apple Worldwide Developers Conference as a Mac OS X game engine. The engine has since been gradually extended to support a variety of desktop, mobile, console and virtual reality platforms. Unity has many features that allow researchers to easily play with them and help make the visualisation more flexible. Unity has the following features:1- Unity is a free, flexible, and accessible platform for simple 2D or very complicated visualisation. 2-In Unity, you can peruse an enormous amount of genres, subgenres, and styles. 3-Unity's realism capabilities are so powerful that many developers use it for other tasks than building games. 4-Unity supports two common programming languages: C$\sharp$ and JavaScript. Anyone with a C$\sharp$ background can quickly jump into Unity and start scripting. 5-One thing about programming in Unity is that you can use the UI intuitively to add functionality to your scripts, versus having to interact with a specific list of variables. 6-Unity has always prioritized building for any platform, and the selection (for iOS, Android, PC, or consoles) is constantly expanding. Unity has support for the Oculus Rift, HTC Vive, Microsoft Hololens, and more. 6- Getting assets is easy. Whatever Unity doesn't have built-in; it can be found in the Asset Store. 7- For future needs Unity Services. This is a new set of features that make it easier for you to build, share, and sell your project. Unity Cloud Build and Unity Collaborate are tools for backing up your entire project and building multiple versions without bogging down your system.}. 

\subsection{Argument Analysis Methods}
Several AM systems include analytical methods available to classify, compare, and measure the argument structure, including \textbf{non-numerical analysis} for argumentative users. For instance: Neva \cite{YGR20}, Carneades \cite{Walton}, and data-set of claims by \cite{durmus2019}; for general users, \cite{Kiesel20}, Kialo \cite{Beck2018} and PEOPLES \cite{Iwan20};  designed to help students and instructors \cite{cullen18}, AGORA-net \cite{Hoffmann2014}, and F2F classroom \cite{HAN20}. Alternatively, systems are based on argument schemes \cite{Hoffmann2018}, Carneades \cite{Gordon07} and Parmenides \cite{Cartwright2009}. 

\textbf{Argument structure numerical analysis}, comparing the argument structure based on the weight of the pros and cons, such as \cite{Brenneis2020} and designVUE \cite{Baroni2015}.  However, these evaluations are complex and lack assessment of argument components, with no supporting intrinsic or extrinsic analysis evaluation.  In contrast, this study assesses argument components and uses a simple numerical process (explicit functionality for mapping decisions) based on positive and negative weights.

\subsection{Evaluation of the Argument Structure Representation}
The study's central part evaluates two representations of arguments: a text-based argument and a 2D graph-based argument based on Walton's argument structure. Several previous studies have conducted a similar evaluation process in varying ways: the findings in \cite{cullen18} deepen understanding of how visualisations support logical reasoning. Alternatively, \cite{Rieche18}'s qualitative and quantitative evaluation of two visual representations, PCM and HTNM. Alternatively, a case study by \cite{Al-Shehhi2015} that includes four systems were implemented based on two cognitive thinking modes. Furthermore, \cite{Riechert18} evaluate the presentation of reasoning in five different visualisation forms. Alternatively, AVISE \cite{Green2019} allows an evaluation of real-world arguments. None of these studies are evaluated based on Walton's argument structure, and they do not support intrinsic or extrinsic evaluation or explicit functionality for mapping decisions. In contrast, the study aims to show the significant differences and impacts of the 2D graph-based argument compared to the text-based argument on argument interpretation, agreement evaluation, performance, working memory, user perception, and user experience. The study assesses the arguments using a visual scale tool (as intrinsic evaluation) and assesses argument models using two questionnaires, NASA TLX and UEQ (as extrinsic evaluation). 

\section{Methods}
\subsection{Development of a 2D Argumentation Graphs (AGs) Model- 2D AGM.}
The target model is the 2D AGM, which requires the following three dimensions.

\textbf{Review of the argumentation mining, corpus selection and annotation.}
Each argument text is considered a debate with two sides—pros and cons—with their respective claims and premises. The text is organised so that it begins with the main conclusions, and then each pro statement is followed by its con statement; this structure is followed for all the argument topics. Unstructured argument text was randomly selected\footnote{} and extracted from debate topics. The selected structure underlying the argumentation mining schema is based on Walton’s classification scheme \cite{Walton2016}, which was revised and simplified by Milz\cite{Milz2017}. We performed configurations on the Brat tool that define entities (claims, premises, and conclusions), relationships (support, attack, undercut, and rebuttal), and attributes (analogy, positiontoknow, expertopinion, popularopinion, causetoeffect, falsification, positiveconsequences, and negativeconsequences), which are based on argument schemes. The argument schemes used are argument from expert opinion; argument from position to know; argument from popular opinion; argument from analogy; argument from positive/negative consequences; argument from cause to effect; and argument from falsification. Indeed, each topic has two conclusions, one pro and one con. As a result, we annotated and linked all pro- and con-side claims, as well as their premises, to the conclusions on both sides. Each premise’s node is annotated, connected with its own attribute (scheme type) based on Walton’s classification schemes, and linked to its claim/claims. The criteria for each scheme have been described by Walton \cite{Walton2008}. This resulted in an annotated corpus, which was parsed to build the AG, (see Fig. \ref{fig:experiment_workflow}(left)).
 \newline

\begin{figure*}[h!]
           \centering
             \begin{subfigure}{\linewidth}
               \centering
               \fbox{\includegraphics[width=0.95\textwidth]{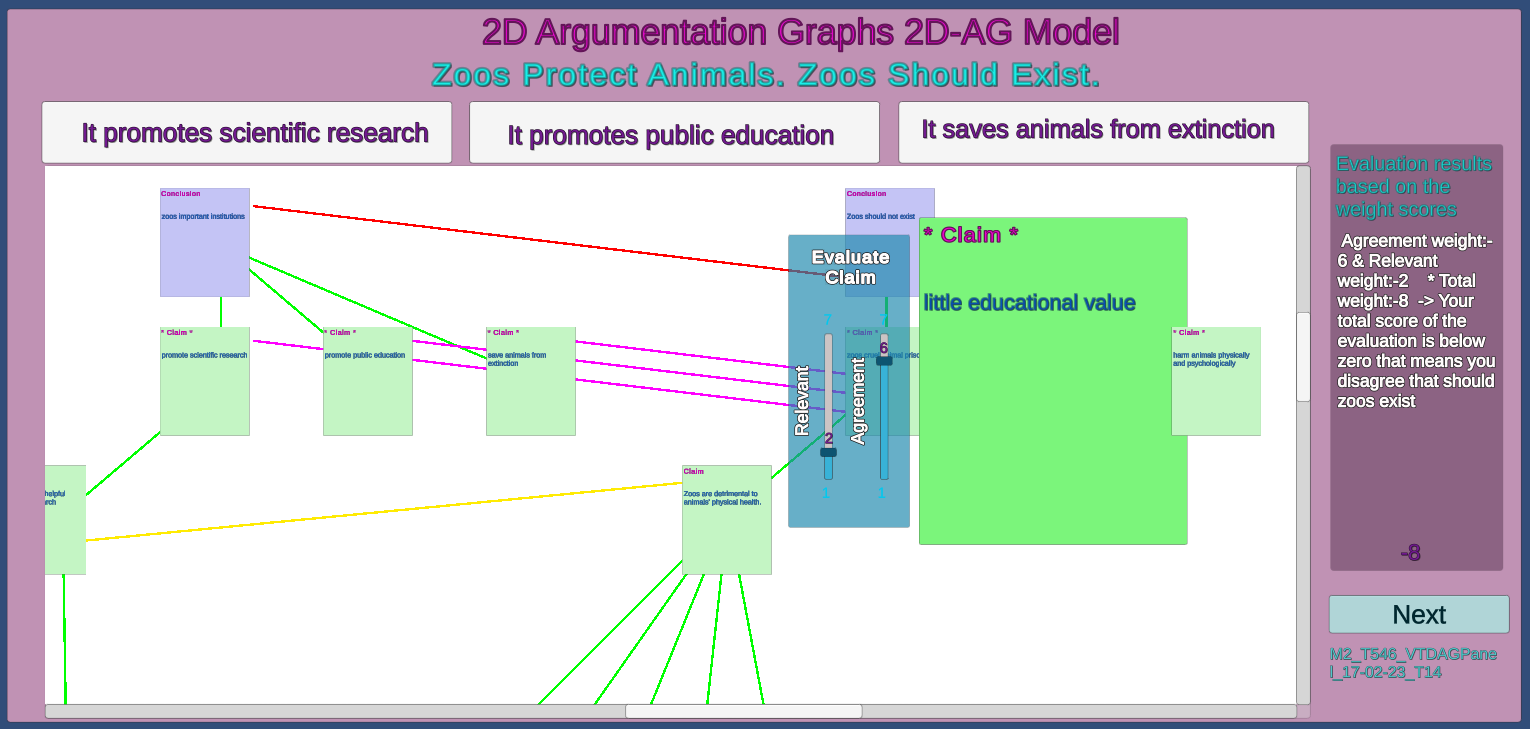}}
                 \caption{A claim (green node) in 2D AGM, which is associated with two Likert-scale sliders for relevance and agreement evaluation (1 to 7), where 1 = disagrees/not relevant, and 7 = strongly agrees/relevant. }
                 \label{fig:2Dmodel_claims}
             \end{subfigure}
              \begin{subfigure}{\linewidth}
                \centering
                \fbox{\includegraphics[width=0.95\textwidth]{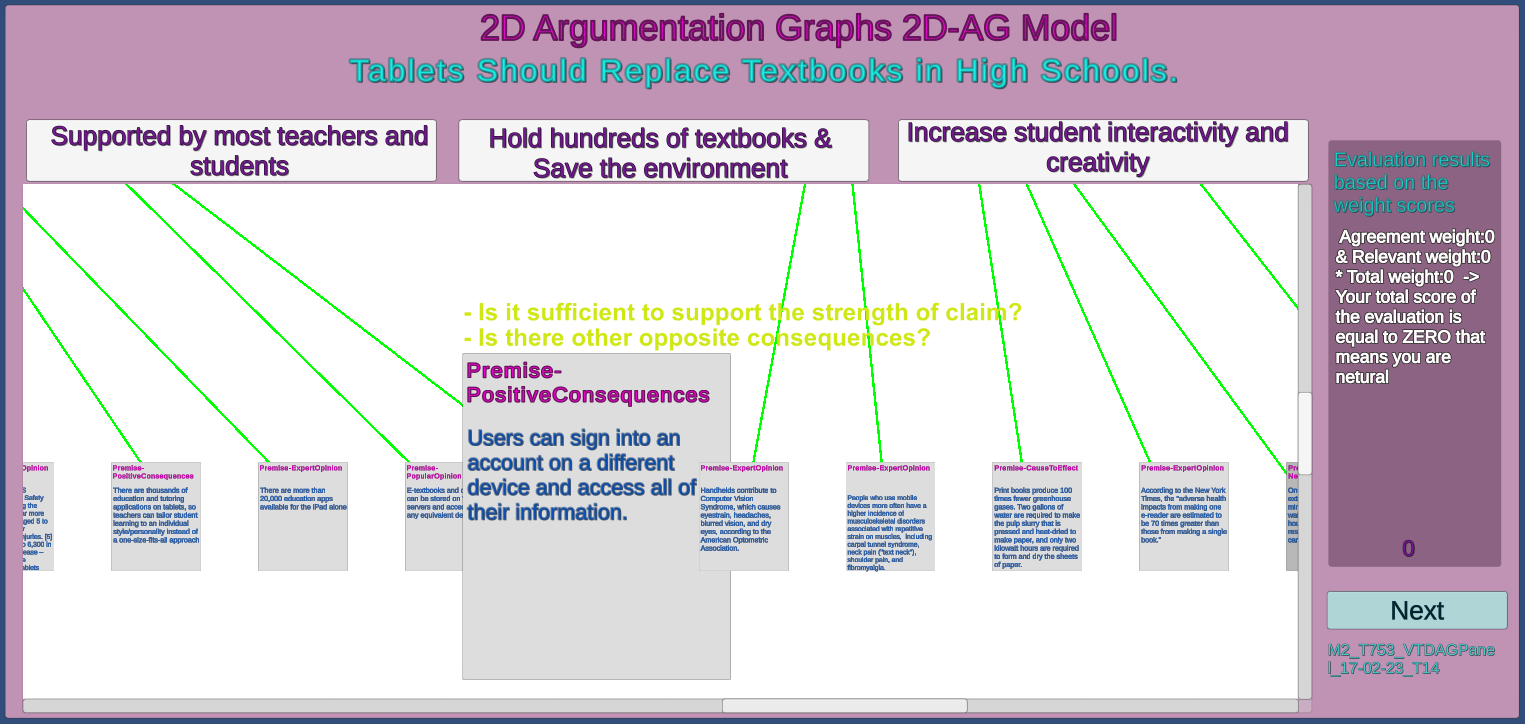}}
                 \caption{A premise (grey node) in 2D AGM, which is associated with an appropriate scheme type and related CQs. The argument scheme types are, for example, an argument from expert opinion, an argument from analogy, etc., etc.\cite{Walton2016}.}
                 \label{fig:2Dmodel_premises}
             \end{subfigure}
         \caption{The 2D Argumentation Graph Model (2D AGM) includes the AG panel (left) that contains three or two main claims (tabs). Each tab contains two AG subtrees: the pro-AG subtree and its counter-con-AG subtree. A coloured rectangular shape represents a node, either a conclusion (purple), a claim (green), or a premise (grey) node. Each coloured link represents the relationship between the nodes, and these either support (green), undercut (pink), rebut (yellow), or attack (red). The right side includes the evaluation message for the overall weights of the relevant agreement scores.}
         
         \label{fig:2Dmodel}
       \end{figure*}

\textbf{Building AG structure.}
These argument structures are mapped onto a tree-based structure \cite{Walton2008} to represent the arguments \cite{Katifori2007, Bembenik2016}. In particular, the tree-based structure uses a non-space-filling variation that can be a hierarchical relationship and is represented in a node-link diagram. A node-link diagram is influenced by two factors: an-out and depth, which makes it easier to render a tree that is not excessively deep. The study employed a node-link diagram to build a 2D AGM with some of the categories and relations of the AG structure. Therefore, each line in an annotated corpus provides either a node, a link, or an attribute associated with the type. A coloured rectangular shape represents a node, either a conclusion (purple), a claim (green), or a premise (grey) node. Each premise node is associated with an (attribute) argument scheme (e.g., an argument from expert opinion, an argument from analogy, etc.)\cite{Walton2016}. Bright colour nodes mean that they have not been clicked yet, and dark colour means that the participant has clicked on the node. Each coloured link represents the relationship between the nodes, and these either support (green), undercut (pink), rebut (yellow), or attack (red). In an AG, spatial and colour configurations vary with the supporting categories, (see Figure \ref{fig:2Dmodel}). 
     
\textbf{Rendering AG structure in 2D AG model}.\label{agm score_aggregation}
In Unity, we used a rectangular panel, a common rendering tool, to present an AG on the screen. On the left-hand side is the AG panel, which contains three or two main claims (tabs). Each tab contains two AG subtrees: the pro-AG subtree and its counter con-AG subtree. The AG sub-structure is rendered inside an asset called ‘Scroll View’ to allow a user to explore the graph by scrolling up and down, and/or left and right. For each \textbf{claim node} (green node) two Likert-scale sliders are used (for relevance and agreement), measuring the degree to which a claim is relevant to the conclusion, and to what degree a user agrees with that claim. The score is from 1 to 7, where 1 = disagrees/not relevant, and 7 = strongly agrees/relevant. Each \textbf{premise node} (grey node) is associated with a scheme type and related CQs to test the strength and acceptability of the argument.
On the right-hand side of the rectangular panel is the evaluation message of the whole argument, which is based on the aggregated score of all the claims  (see Fig. \ref{fig:Evaluation-result}). Each time a user adds a score to the claim, the result of the evaluation message is automatically updated. As a result, an arithmetic summation of the scores from all the claims gives the final decision of that argument. Additionally, there is a hidden timer to compute the total time spent by a user on the model. 

\subsection{Development of Pure Text-based Model-PTM.} \label{ptm des and scores}
The baseline model, pure text-based argumentation (PTM), is delivered by a simple renderisation of the text (under the same set of debate topics), without the additional structure.

\begin{figure}[!ht]
        \centering
        \fbox{\includegraphics[width=0.95\textwidth]{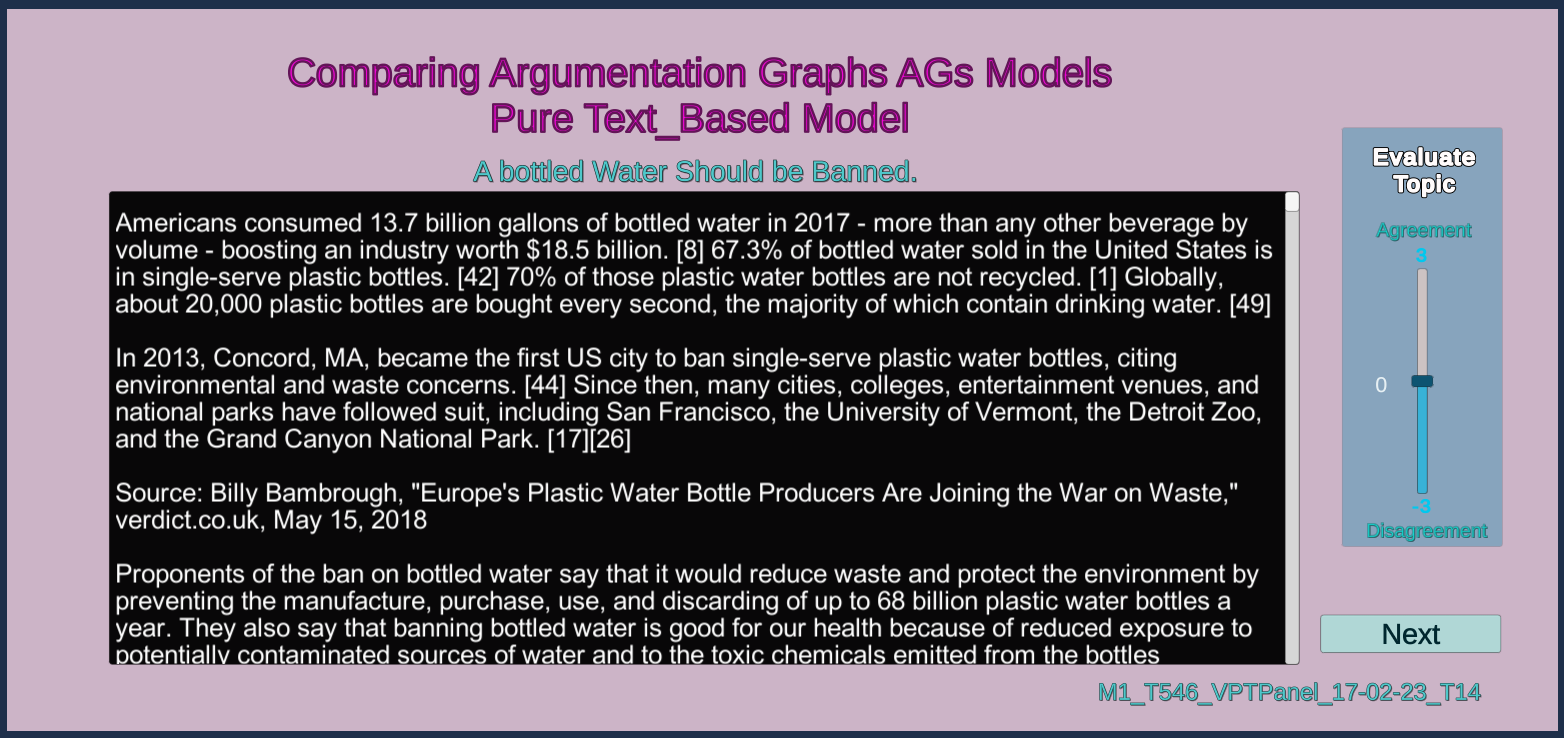}}
        \caption{Pure text-based model (PTM, baseline model), which includes the argument text on the left-hand side. On the right, the user moves a Likert-scale slider to indicate their agreement with an argument (-3 to +3), where -3 indicates disagreement, +3 indicates agreement, and 0 means neutral. }
       
        \label{fig:PureText-based-Model}
    \end{figure}

In Unity, we used a rectangular panel to present the argument text on the screen. A component called `TextMeshPro' inside another component called `Scroll View' is used within the rectangular panel to allow users to scroll over the text. The text is rendered one page long on a dark background for easier reading. Additionally, there is a Likert-scale slider on the left with which the users can evaluate an argument on a scale from -3 to +3, where -3 indicates disagreement, +3 indicates agreement, and 0 means neutral. There is also a hidden timer to calculate the total time spent by a user  (see Fig. \ref{fig:PureText-based-Model}).

    \section{The Experimental Design}
The experiment was conducted with a within-subject design. Each participant is presented with two models: first PTM, and then 2D AGM; for each participant, two arguments are randomly selected from a set of 10 arguments (each from a different topic, see Supp \ref{argument topics}).

\begin{figure*}[!ht]
        \centering
        \begin{subfigure}{\linewidth}
            \centering
            \fbox{\includegraphics[width=0.9\textwidth]{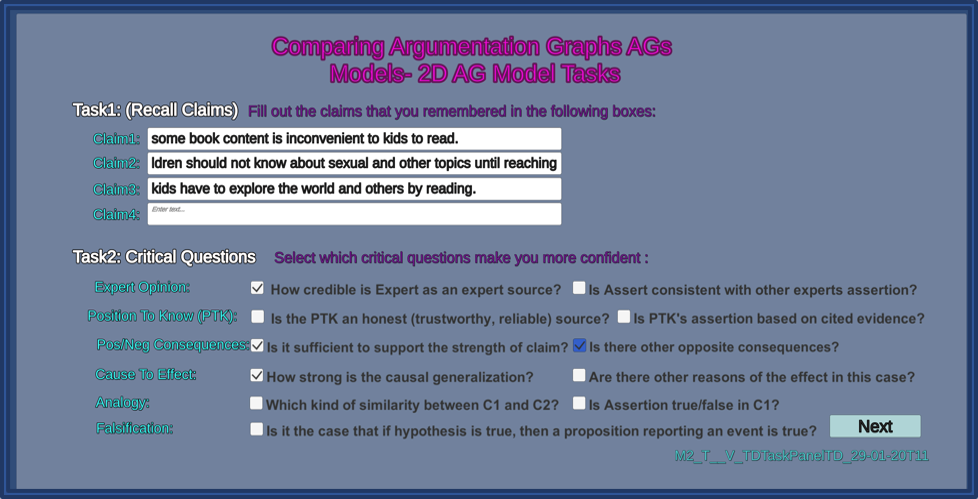}}
            \caption{2D AGM tasks - Task 1: Recall Claims includes four input fields to enter the recall claims. Task 2: Select CQs for the 2D AGM Model}
            \label{fig:task12}
        \end{subfigure}
        \begin{subfigure}{\linewidth}
            \centering
            \fbox{\includegraphics[width=0.9\textwidth]{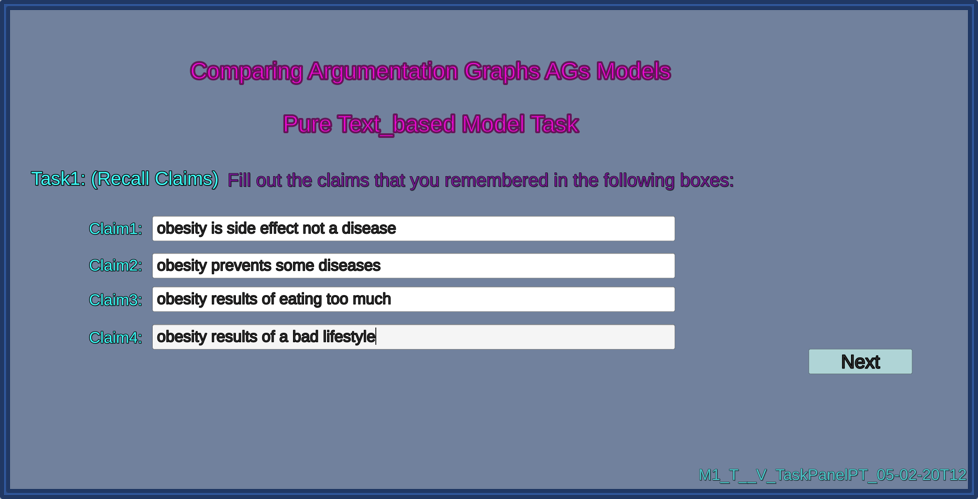}}
            \caption{PTM tasks - Task 1: Recall Claims for the PTM Model includes four input fields to enter the recall claims.}
            \label{fig:task1}
        \end{subfigure}
        \caption{Models tasks- Task 1: Recall Claims for (a) the 2D AGM and (b) PTM Models includes four input fields to enter the recall claims. Task 2:  Select CQs for the 2D AGM Model. }
        \label{fig:alltasks}
\end{figure*}

In the first part of the experiment, the PTM is evaluated. Initially, the user is asked about the topic’s perceived controversy, and their familiarity with and initial position on the debate. Then, the user is presented with an argument essay in the text-based model, presented in the scrolled view text. They indicate their level of agreement with the argument using the Likert-scale controller (Task 1, see Fig. \ref{fig:PureText-based-Model}). After that, the user is asked about the topic’s strength, and whether they liked the topic and liked the model. Then, the user is asked to write the memorised claims by filling in four text-field boxes with the claim statements (later referred to as Task 1: Recall Claims, see Fig. \ref{fig:task1}). Finally, the user fills in two short close-ended structured questionnaires: NASA-TLX \ref{experiment questionnaires}.\ref{nasa tlx sec} and UEQ \ref{experiment questionnaires}.\ref{ueq_sec} to evaluate the workload, and their experience, and opinions (see Fig. \ref{fig:nasa-tlx}, and \ref{fig:UEQ}).

In the second part, the user interacts with the 2D AGM. Similarly to the first part of the experiment, before the 2D AG is shown, the user is asked about their familiarity with the topic, its controversy level and their initial position. The 2D AG is presented inside the scrolled view as a 2D-coloured tree graph (Fig.\ref{fig:2Dmodel}). The user indicates their agreement with the argument and the relevance of each main claim, utilising the Likert-scale controller attached to each claim. The final score for agreement is produced as an aggregation of all the scores (see Sec.\ref{agm score_aggregation}). 

In the next step, the user is asked to perform two tasks: Task 1: Recall Claims and Task 2: Select CQs. In Task 1, the same as for PTM, the user writes the memorised claims in four text-field boxes with the claim statements. In Task 2, the user is presented with a set of CQs with check boxes (from Walton’s book) and they are asked to click those that were used in the argumentation in the 2D AGM (see Fig. \ref{fig:task12}). Finally, as for the PTM, the user completes two questionnaires: NASA-TLX and UEQ.
For details regarding the experimental environment refer to Supp methods \ref{tasks_methods}, NASA-TLX \ref{experiment questionnaires}.\ref{nasa tlx sec}, UEQ \ref{experiment questionnaires}.\ref{ueq_sec}. For reproducibility purposes, the full experimental workflow is available at Supp Fig.\ref{fig:workflow}. 

\subsection{Participants, Metrics and Settings}
Data: 10 topics representing different arguments (see Supp. Sec.\ref{argument topics} ) were represented in the PTM and the 2D AGM model.

Participants: we recruited 25 postgraduate students from the University of [anonymised] 
        School of Engineering and compensated participants with Amazon gift cards(£10/participant). Fifteen women and 10 men participated, with 23 participants being PhD students and 2 being master's students. The participants' ages ranged from 20 to 40. All of the participants had a normal or corrected-to-normal vision, without a colour deficiency.

Metrics: The experiment aims to evaluate the following aspects:
\begin{enumerate}
         \item The user's initial position in the debate (before seeing PTM or 2D AGM).
         \item Level of agreement with the argument (as they read PTM or 2D AGM), see section \ref{ptm des and scores} and \ref{agm score_aggregation}.  
         \item Time to complete the task (the total time spent on a model in minutes and seconds). 
         \item Task 1: Working memory (short-term memory) (A-BLEU was calculated in decimal numbers between 0 and 1, see Supp. Sec. \ref{tasks_methods}.\ref{task1_bleu_def}).
         \item Task 2: Selected CQs. (the total number of selected CQs in integer number (0-11). see Supp. Sec. \ref{Walton's CQs}, \ref{tasks_methods}.\ref{task2_CQS} and \ref{cqs-sec}) 
         \item Scales of participants' familiarity with the topic(prior knowledge about the topic); controversy(complex and wide-range debate among the public), strength (reasons - premises statements - to support the claims and conclusions); and preferences (liking the topic). Calculate the Likert-Scale controller rating 1-7, where 1 is strongly negative and 7 is strongly positive.
         \item Workload (NASA-TLX) in terms of temporal demand\footnote{Temporal Demand: How much time pressure did you feel due to the rate or pace at which the task elements occurred? Was the pace slow and leisurely or rapid and frantic?}, physical demand\footnote{Physical Demand: How much physical activity was required (e.g., pushing. pulling, turning. etc.)? Was the task easy or demanding? Slow or brisk? Slack or strenuous?}, mental demand\footnote{Mental Demand: How much mental perceptual activity was required (e.g., thinking, searching, remembering. etc.)? Was the task easy or demanding? Simple or complex?}, performance\footnote{Performance: How successful do you think you were in accomplishing the goals of the task set by the experimenter (or yourself)? How satisfied were you with your performance in accomplishing these goals?}, frustration\footnote{Frustration: How insecure. discouraged. irritated stressed or annoyed versus secure. content. relaxed or complacent did you feel during the task?}, and effort\footnote{Effort: How hard did you have to work (mentally and physically) to accomplish your level of performance?}. The classification result for each of them is an integer (0-500) and the total workload is a decimal (0-100).
         \item User experience (UEQ) in terms of attractiveness\footnote{Attractiveness means: Overall impression of the model. Do users like or dislike it?}, perspicuity\footnote{Perspicuity means: Is it easy to become familiar with the model and to learn how to use it?}, efficiency\footnote{Efficiency means: Can users solve their tasks without unnecessary effort? Does it react fast?}, dependability\footnote{Dependability means: Does the user feel in control of the interaction? Is it secure and predictable?}, and stimulation.\footnote{Stimulation means: Is it exciting and motivating to use the model? Is it fun to use?}. The rating result for each of them is a decimal number (-3 to 3).
         \item User experience (UEQ) in terms of benchmark\footnote{Benchmark means that the scale is set in relation to existing values from a benchmark dataset. This set contains data from 21,175 people from 468 studies concerning different products (software, web pages, web shops, etc)}. The expected results are a decimal number (-3 to 3). 
\end{enumerate}
 Setting: The experiment was conducted in a quiet room during the day. The project was run on the Unity game engine as a platform. The participants interacted with a mouse and keyboard to complete the online tasks through MS Teams. The results and data were collected using a programmatic embedded plugin package called ``Uni-Excel", which has excellent benefits and assistance, see Supp \ref{Uni-Excel}.

\begin{table*}
\small
\centering

\caption{Comparisons between the PTM and 2D AGM. Sig - significance: *** for p<0.001,** for p<0.01, * for p<0.05, ns (non significant) for p>0.05; p values from Wilcoxon signed-ranks test, and t-test for UEQ and NASA TLX.}
\label{tab:allResults}
\begin{tabular}{p{2.5 cm} p{4.2 cm} p{2.5 cm} p{2.5cm} p{1 cm} p{0.7 cm}}
                                                                                                        &                                                                                      & \textbf{PTM}                                           & \textbf{2D AGM}                                      & \textbf{p} & \textbf{ Sig}  \\ 
\midrule
Time to interpret an argument~~                                                                         & Median [Q1,Q3]                                                                       & 11:13 [5:57, 17:12] min                                & 13:05 [9:33, 23:13] min                              & 0.015      & *              \\ 
\midrule
\begin{tabular}[c]{@{}l@{}}Task 1\\‘Recall Claims’~~\end{tabular}                                       & \begin{tabular}[c]{@{}l@{}}The A-BLEU Scores\\Median [Q1,Q3]\end{tabular}            & 0.07 [0.02, 0.14]                                      & 0.10 [0.04, 0.2]                                     & 0.247      & ns             \\ 
\midrule
\multirow{7}{*}{\begin{tabular}[c]{@{}l@{}}Opinion of\\participants\\(Likert Scale [1-7])\end{tabular}} & \begin{tabular}[c]{@{}l@{}}Controversial Topics, n\\Median [Q1,Q3]\end{tabular}      & \begin{tabular}[c]{@{}l@{}}17\\5 [3, 6]\end{tabular}   & \begin{tabular}[c]{@{}l@{}}20\\6 [4, 7]\end{tabular} & 0.148      & ns             \\
                                                                                                        & \begin{tabular}[c]{@{}l@{}}Familiarity Topics, n\\Median{[}Q1,Q3]\end{tabular}       & \begin{tabular}[c]{@{}l@{}}16\\5 [3, 6]\end{tabular}   & \begin{tabular}[c]{@{}l@{}}12\\4 [2.5, 5]\end{tabular} & 0.33       & ns              \\
                                                                                                        & \begin{tabular}[c]{@{}l@{}}Strong Topics, n\\Median [Q1,Q3]\end{tabular}             & \begin{tabular}[c]{@{}l@{}}22\\6 [5, 7]\end{tabular}   & \begin{tabular}[c]{@{}l@{}}23\\6 [5, 7]\end{tabular} & 0.84       & ns             \\
                                                                                                        & \begin{tabular}[c]{@{}l@{}}Liked Topic, n\\Median [Q1,Q3]\end{tabular}               & \begin{tabular}[c]{@{}l@{}}17\\6 [4, 6.5]\end{tabular} & \begin{tabular}[c]{@{}l@{}}19\\6 [4.5, 7]\end{tabular} & 0.3        & ns             \\
                                                                                                        & \begin{tabular}[c]{@{}l@{}}Liked Model, n\\Median [Q1,Q3]\end{tabular}               & \begin{tabular}[c]{@{}l@{}}9\\3 [2, 5]\end{tabular}    & \begin{tabular}[c]{@{}l@{}}21\\6 [4, 7]\end{tabular} & 0.005      & **             \\
                                
                                      & Initial Position Change, n                                                           & 10                                                     & 13                                                   & -      & -          \\
                                                                                                        & \begin{tabular}[c]{@{}l@{}}Absolute magnitude of change\\Median [Q1,Q3]\end{tabular} & 3 [2.75, 4]                                               & 2 [1.5, 3]                                             & 0.29       & ns             \\ 
\hline
\multirow{5}{*}{\begin{tabular}[c]{@{}l@{}}UEQ\\(Rating [-3,3])\end{tabular}}                           & Attractiveness , Median [Q1,Q3]                                                      & -0.5 [-1.25, 1.5]                                      & 1 [0.63, 1.75]                                       & 0.0026      & **             \\
                                                                                                        & Perspicuity, Median [Q1,Q3]                                                          & 1 [0, 1.67]                                            & 1.33 [0.83, 1.67]                                       & 0.059      & ns             \\
                                                                                                        & Efficiency, Median [Q1,Q3]                                                           & 0.5 [-0.5, 2]                                             & 1.5 [0.75, 2.25]                                        & 0.036      & *             \\
                                                                                                        & Dependability, Median [Q1,Q3]                                                        & 0 [-0.5, 1.5]                                          & 1.5 [0, 1.75]                                        & 0.113      & ns             \\
                                                                                                        & Stimulation, Median [Q1,Q3]                                                          & 0.5 [-0.25, 0.88]                                     & 0.75 [0.5, 1.13]                                    & 0.035      & *              \\ 
\midrule
\multirow{6}{*}{\begin{tabular}[c]{@{}l@{}}NASA TLX\\(Rating [0-500])\end{tabular}}                     & Mental Demand, Median [Q1,Q3]                                                        & 300 [210, 387.5]                                       & 120 [65, 205]                                        & 0.0002      & ***            \\
                                                                                                        & Physical Demand, Median [Q1,Q3]                                                      & 0 [0, 60]                                              & 0 [0, 95]                                            & 0.50       & ns             \\
                                                                                                        & Temporal Demand, Median [Q1,Q3]                                                      & 100 [50, 180]                                          & 100 [75, 210]                                        & 0.32       & ns             \\
                                                                                                        & Performance, Median [Q1,Q3]                                                          & 100 [65, 232.5]                                        & 130 [27.5, 232.5]                                      & 0.91       & ns             \\
                                                                                                        & Effort, Median [Q1,Q3]                                                               & 200 [90, 300]                                         & 140 [72.5, 210]                                        & 0.07       & ns             \\
                                                                                                        & Frustration, Median [Q1,Q3]                                                          & 120 [7.5, 197.5]                                        & 65 [37.5, 167.5]                                       & 0.51       & ns             \\ 
\midrule
\begin{tabular}[c]{@{}l@{}}NASA TLX\\(Rating [0-100])\end{tabular}                                      & Overall Workload Scores, Median [Q1,Q3]                                              & 64.33 [53.67, 69.5]                                    & 48.33 [39.167, 62.67]                                & 0.01       & *   \\  \bottomrule
\end{tabular}

\end{table*}

\section{Results}
\subsection{Opinion and impressions from interacting with the two models. 2D AGM was preferred by a greater number of participants}
\label{opinion_analysis} 
\begin{figure*}[!ht]
                \centering 
                \fbox{\includegraphics[width=0.55\textwidth] {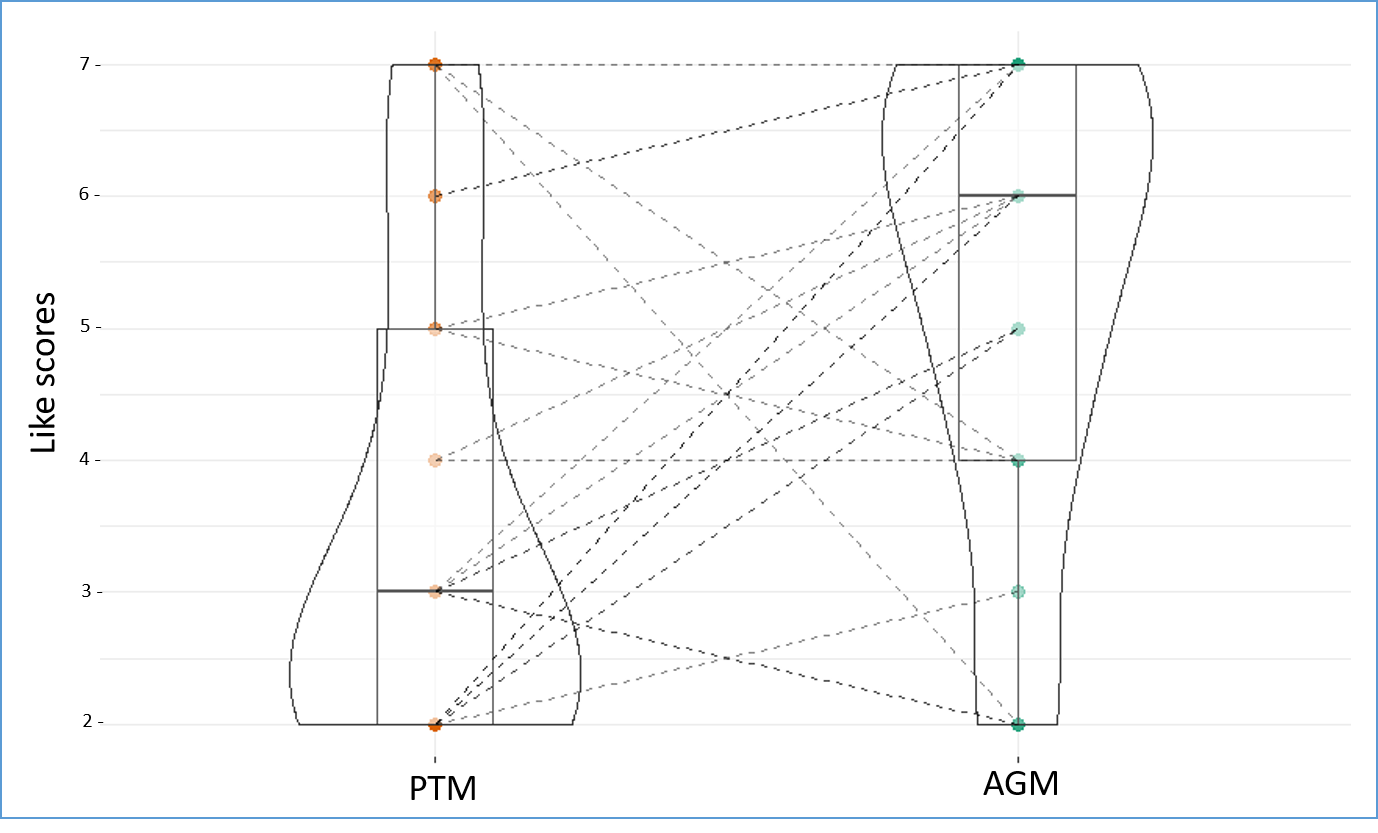}}
                 \caption{The paired plot of the Like model scores for both models is based on the Likert-scale value (>3). Differences measured using p-value from Wilcoxon signed-ranks test, Sig – significance: *** for p < 0.001,** for p < 0.01, * for p < 0.05, ns (non-significant) for p > 0.05) }
                 
                 \label{fig:like-pairedplot}
         \end{figure*}

 As shown in Figure \ref{fig:like-pairedplot}, even though PTM was preferred by some of the participants (N = 9, 36\%), 2D AGM was preferred by a greater number of participants (N = 21, 84\%) - assessment based on the Likert-scale value (>3). The 2D AGM achieved significantly higher scores for being preferred by participants  compared to PTM(p = 0.005, Wilcoxon signed-ranks test). This result proved that 2D AGM has more influence in terms of the set of target metrics and user perception, which are considered parts of \hyperref[RQ4]{RQ4} and \hyperref[RQ5]{RQ5}. Similarly, this reinforces a similar finding from \cite{Riechert2018} for the HTNM model.
 
  \subsection{ Changed decision and agreement: No significant difference between PTM and 2D AGM} \label{change decision}
\begin{figure*}[!ht]
                \centering
                \begin{subfigure}{.5\linewidth}
                  \centering
                  \fbox{\includegraphics[width=0.98\textwidth]{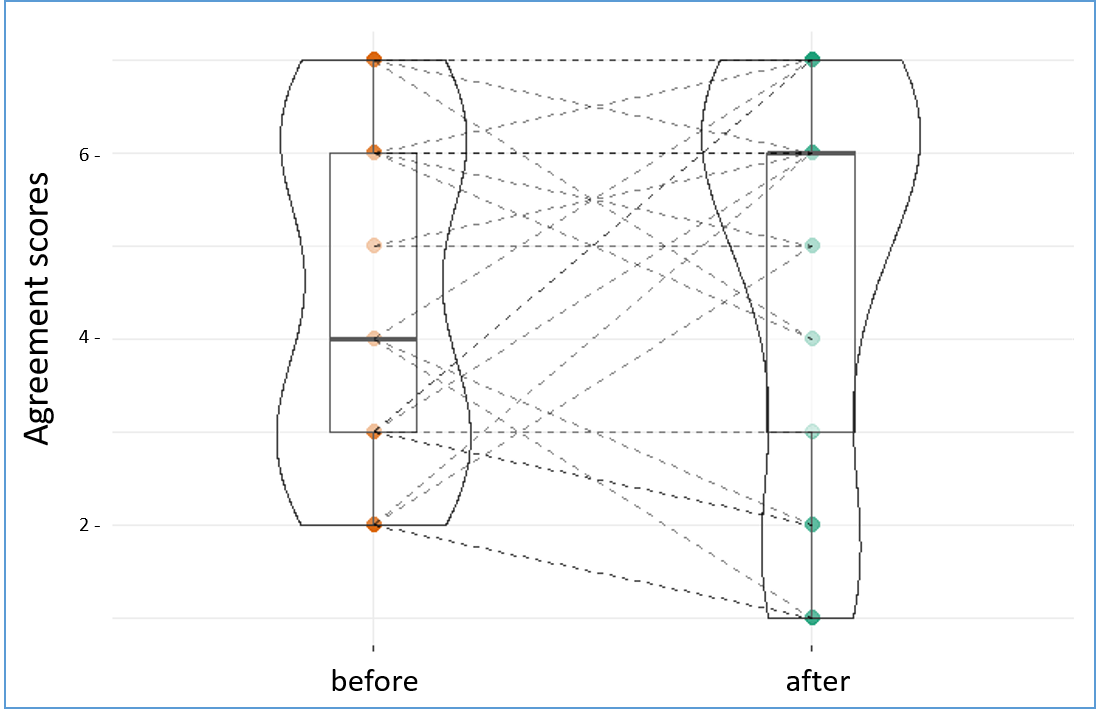}}
                  \caption{PTM.}
                  \label{fig:ptm_cd}
                \end{subfigure}%
                \begin{subfigure}{.5\linewidth}
                  \centering
                  \fbox{\includegraphics[width=0.98\textwidth]{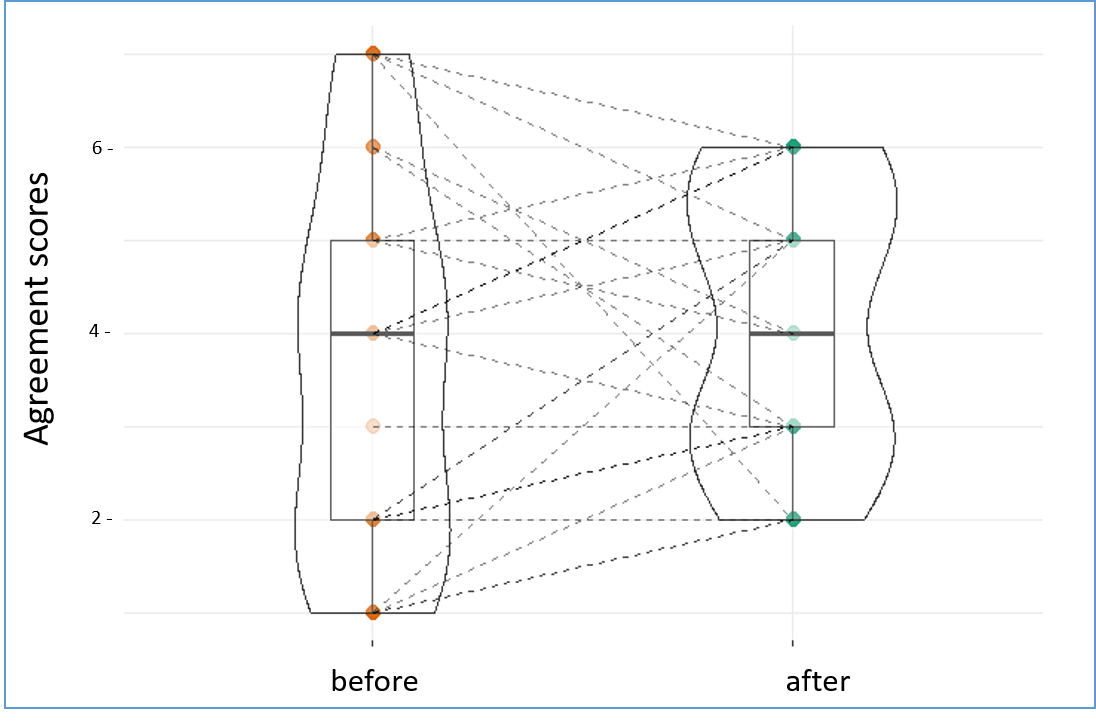}}
                  \caption{2D AGM.}
                  \label{fig:agm_cd}
                \end{subfigure}
                \caption{Agreement scores before and after interaction with two models: (a) PTM, and (b) 2D AGM. Dashed lines connect scores for individual participants. }
               
                \label{fig:Agreement_scores_plot}
         \end{figure*}

     We evaluated whether the participants changed their position after being presented with the model (from the agreement (5–7) to disagreement (1–3) or neutral (4) and vice versa). In the PTM, 10 (40\%) participants changed their position, compared to 13 (52\%) in the 2D AGM. The change was defined in both directions (pro and con): 6 positive changes for PTM and 8 positive changes for 2D AGM, 4 negative changes for PTM and 5 for 2D AGM (see Fig. \ref{fig:Agreement_scores_plot}). However, this difference was not significant when comparing the values of the change between models (p=0.29, see Table \ref{tab:allResults}).
   Although these differences were marginally significant, they still show that 2D AGM supports argument interpretation and construction \hyperref[RQ1]{RQ1} and \hyperref[RQ3]{RQ3}, has a positive effect on the interpretation process \hyperref[RQ2]{RQ2}, and perform better for argument agreement evaluation in terms of the number of the participants who changed their position (intrinsic metrics) compared to PTM \hyperref[RQ4]{RQ4}. Although some other studies, such as\cite{Brenneis2020} and designVUE \cite{Baroni2015}, compare the argument structure based on the weighting of the pros and cons, they carry out a complex and challenging numerical evaluation of the argumentative process, with no supporting argument agreement evaluation or argument model evaluation. In contrast, 2D AGM assesses the argument components and applies a simple numerical process (explicit functionality for mapping decisions) based on positive and negative weights.

    \subsection{Task 1 Recall Claims: Measuring the working memory of participants. 2D AGM and PTM have similar effects in relation to working memory.} \label{BLEU_analysis}
        We compared the effect of the models on the working memory of participants via the A-BLUE score (see Table \ref{tab:allResults}). Both models achieved low values of A-BLUE with no significant differences (p = 0.247), PTM median = 0.07, and 2D AGM median = 0.10. However, we argue that for 2D AGM the A-BLUE score can be improved with more interaction, practice and familiarity with the design configuration. Therefore, the results support \hyperref[RQ1]{RQ1},\hyperref[RQ2]{RQ2}, and \hyperref[RQ3]{RQ3} in terms of effecting working memory positively and supporting the interpretation process. \\ This result has been reported in similar studies such as \cite{cullen18}, which found that organizing good pedagogical practices around collaborative argument visualization leads to meaningful improvements in students’analytical-reasoning skills; also, \cite{Al-Shehhi2015} found that narrative and graphical representation had no effect on the participants’ performance in terms of constructing knowledge. For additional information,  Table \ref{tab:reclaims_sentences} presents the candidates for both models    that were awarded the maximum A-BLEU score values – one participant in PTM and one in 2D AGM. Table \ref{tab:PTM_claims} contains the references for PTM model and Table \ref{tab:AGM_claims} contains the references for the 2D AGM model. The A-BLEU scores for all participants are shown in Fig. \ref{fig:bleu-results-all}.

           \begin{figure*}[!ht]
             \centering
             \fbox{\includegraphics[width=0.55\textwidth]{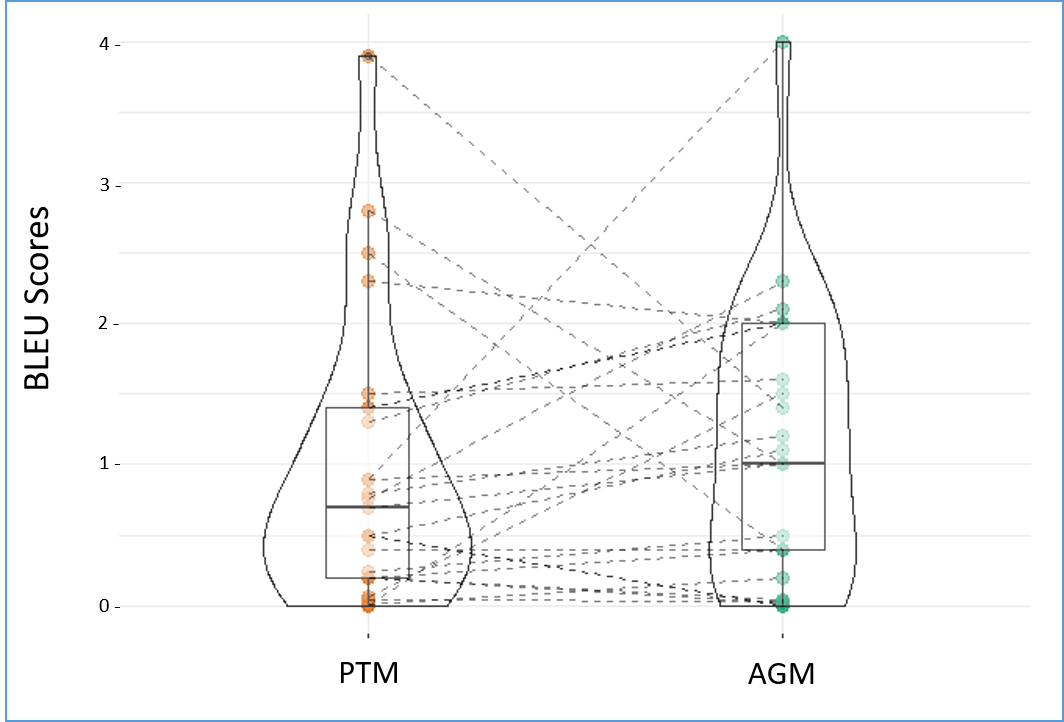}}
             \caption{The paired plot of A-BLEU scores in Task 1 `Recall Claims' for both models. Dashed lines connect scores for individual participants. No significant difference (p=0.247) between the scores achieved by participants in Task 1 in PTM and 2D AGM.  }
            
             \label{fig:BLEU-pairedplot}
          \end{figure*}
  
         \subsection{2D AGM-Task 2: \large{Number of selected CQs correlated with topic's familiarity}}
         \label{CQs_analysis}
\begin{figure*}[!ht]
                \centering
                \begin{subfigure}{.5\linewidth}
                  \centering
                  \fbox{\includegraphics[width=0.90\textwidth]{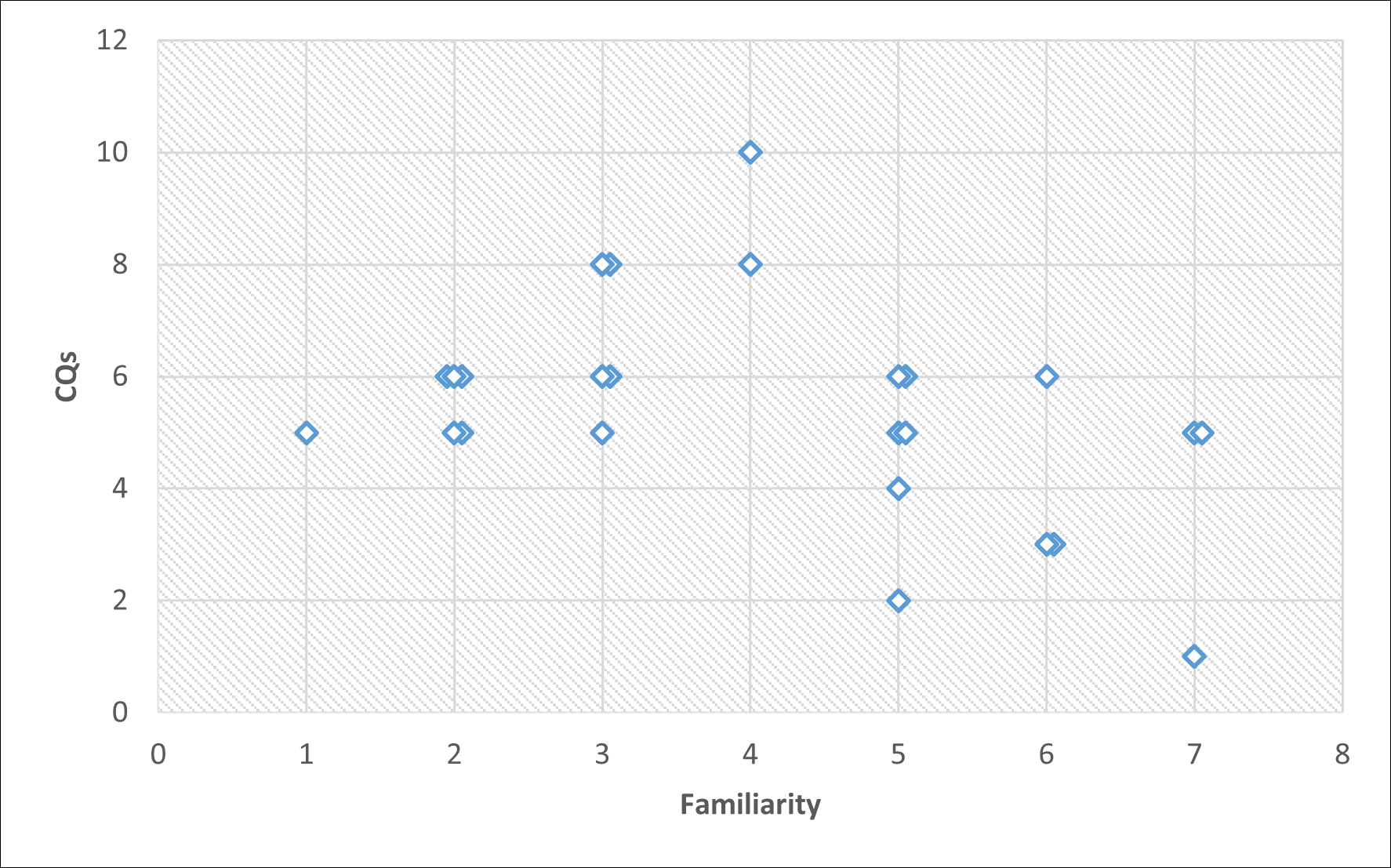}}
                  \caption{Familiarity with the topic vs. no. CQs}
                  \label{fig:linear_fami_cqs}
                \end{subfigure}%
                \begin{subfigure}{.5\linewidth}
                  \centering
                  \fbox{\includegraphics[width=0.90\textwidth]{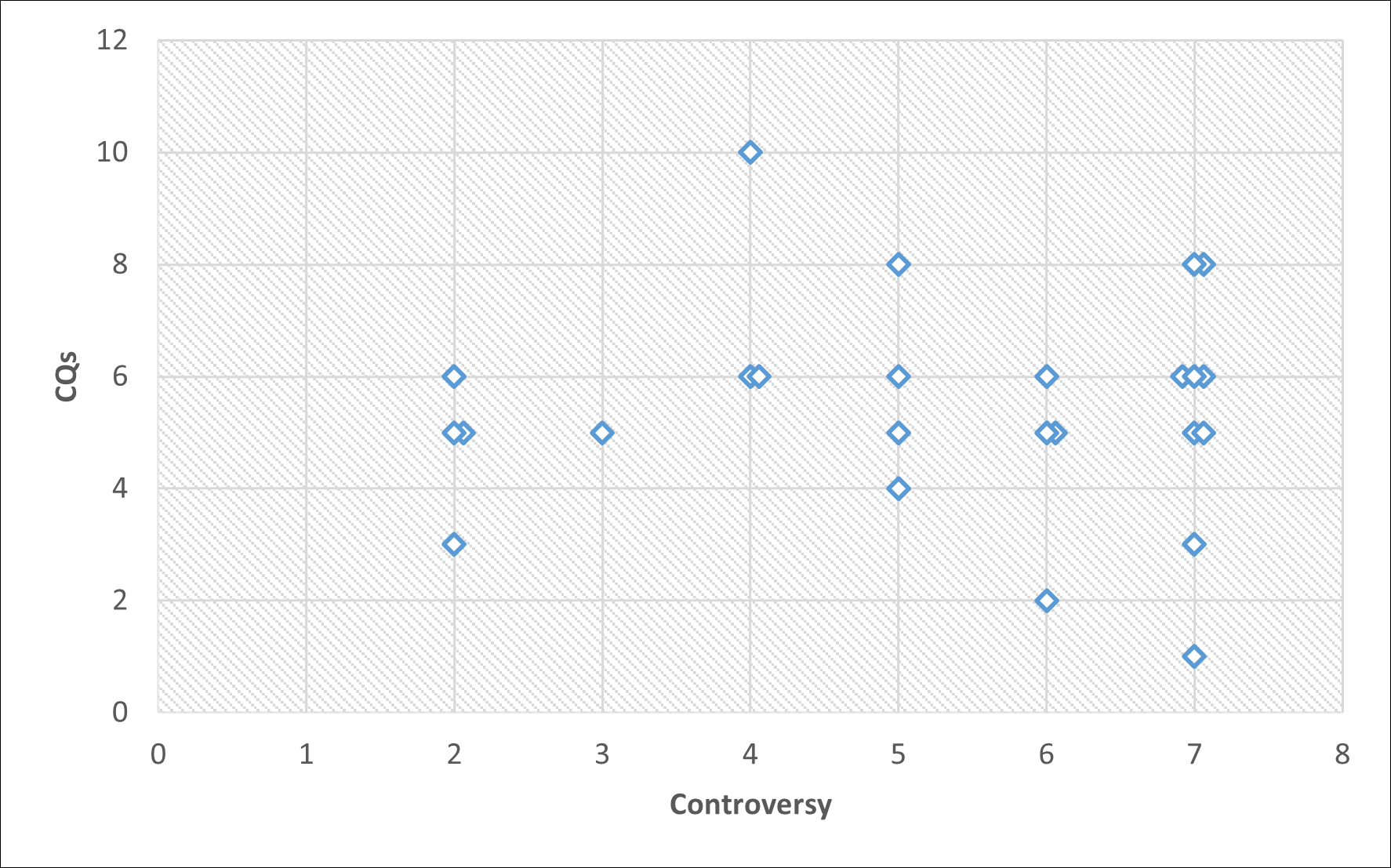}}
                  \caption{Controversy vs. no. CQs}
                  \label{fig:linear_contr_cqs}
                \end{subfigure}
              
                \begin{subfigure}{.5\linewidth}
                  \centering
                  \fbox{\includegraphics[width=0.90\textwidth]{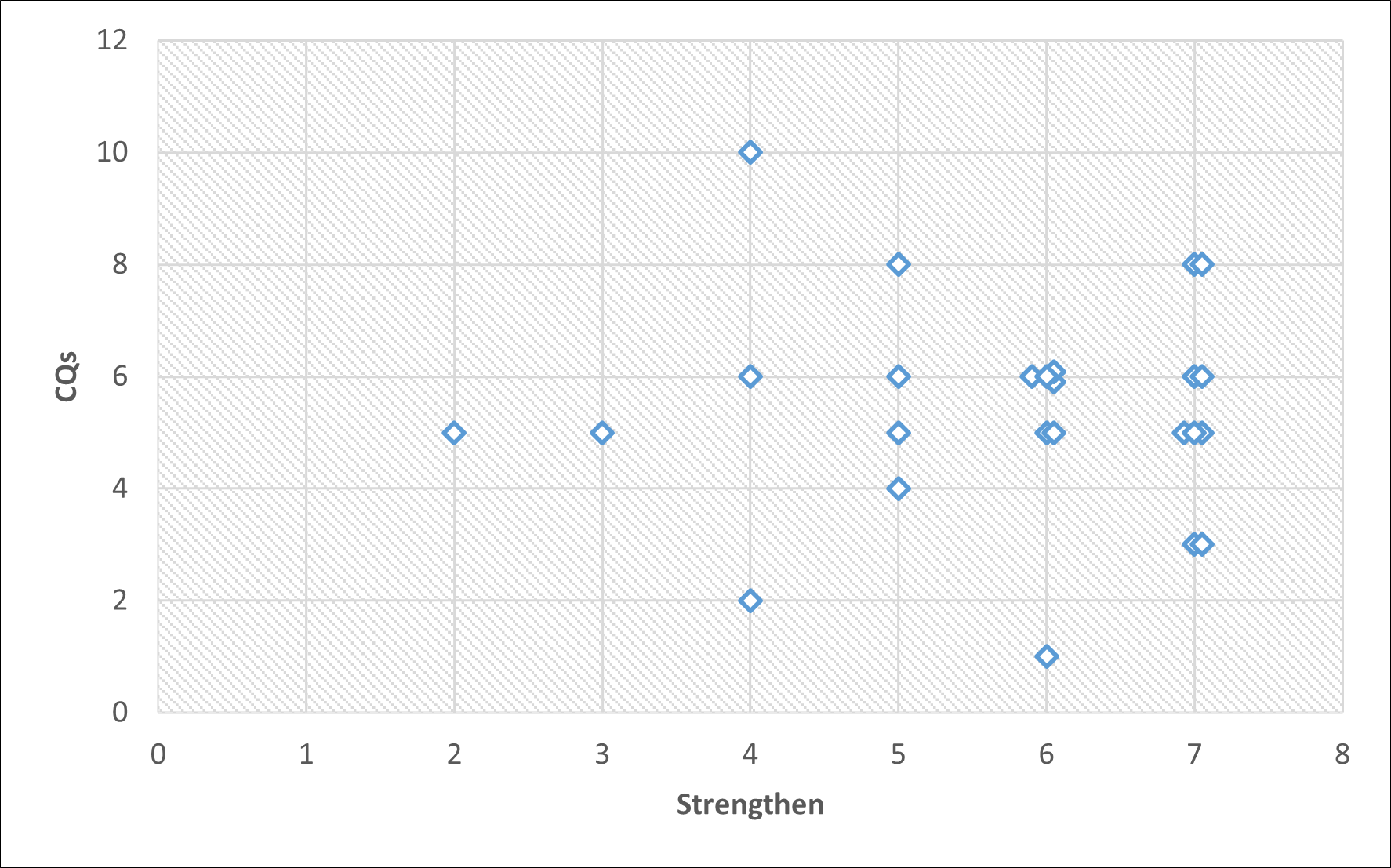}}
                  \caption{Strength of the topic vs. no. CQs}
                  \label{fig:linear_Str_cqs}
                \end{subfigure}%
                \begin{subfigure}{.5\linewidth}
                  \centering
                  \fbox{\includegraphics[width=0.90\textwidth]{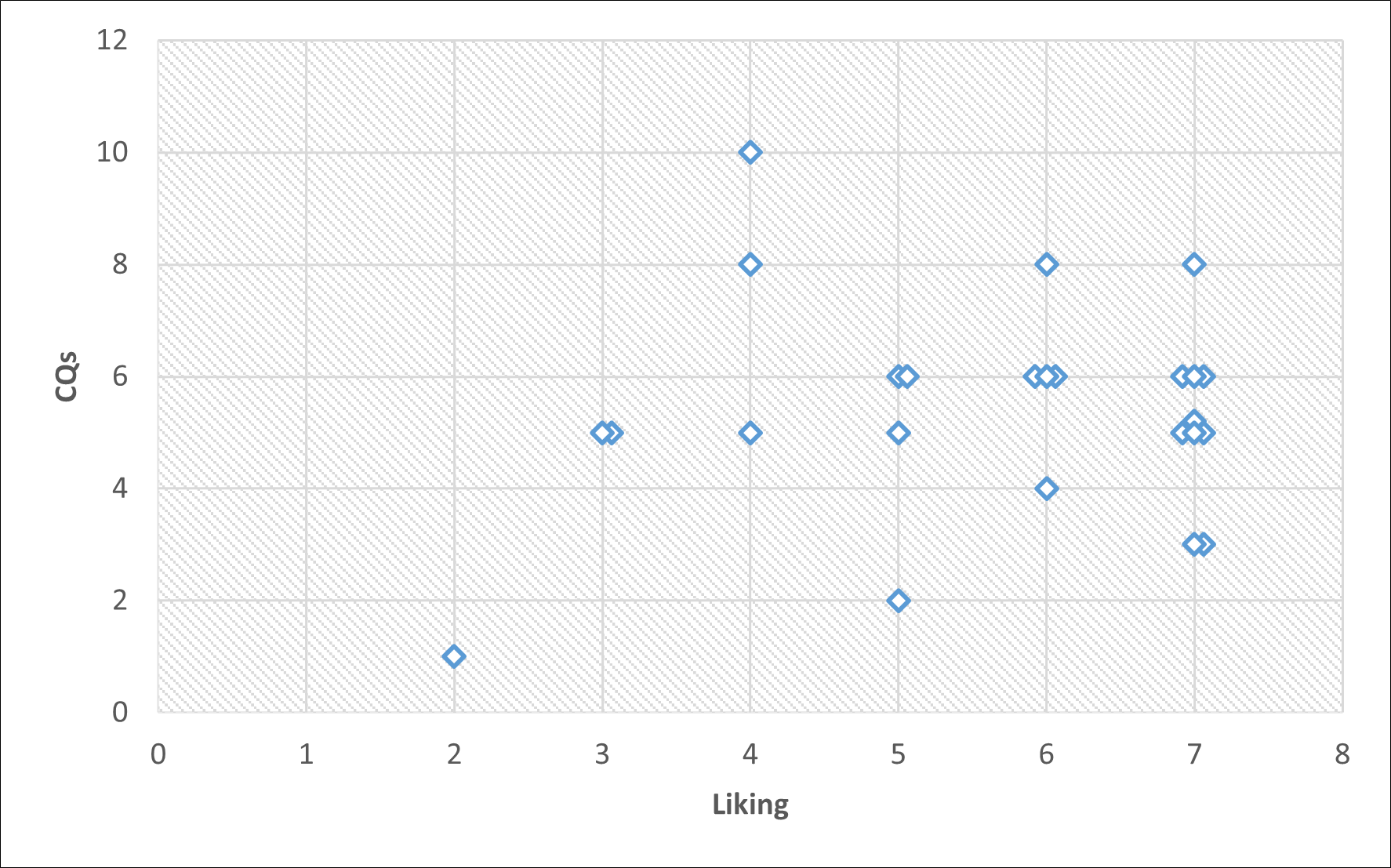}}
                  \caption{Liking the topic vs. no. CQs}
                  \label{fig:linear_like_cqs}
                \end{subfigure}
                \caption{The linear regression of the topic's scales: (a) Familiarity, (b) Controversy, (c) Strength, and (d) Liking vs. the number of selected CQs. Each dot represents a single answer from a single participant. There is a negative correlation between familiarity and the number of CQs (p = 0.038, r = -0.42) but no other correlations.
                }
               
                \label{fig:linear4}
         \end{figure*}

          We found a negative correlation between the number of selected CQs and familiarity with the topic (measured using spearman correlation, p = 0.038, r = -0.42, see Figure \ref{fig:linear_fami_cqs}). The more familiar topic, the fewer CQs are selected, which relates to  \hyperref[RQ1]{RQ1},\hyperref[RQ2]{RQ2}, \hyperref[RQ3]{RQ3}, and \hyperref[RQ6]{RQ6}. This indicates that the scheme types (including CQs) help participants to be more critical of their interpretation, especially with familiar topics. However, there are no other significant correlations between topic strength and number of CQs (p = 0.99, Figure \ref{fig:linear_Str_cqs}), position preference (liking) and number of CQs (p = 0.93, Figure \ref{fig:linear_like_cqs}), or topic controversy and number of CQs (p = 0.87, Figure \ref{fig:linear_contr_cqs}). 
 
          \subsection{The User Experience Questionnaire UEQ (Measuring the users' perceptions): The 2D AGM significantly outperforms PTM for attractiveness, perspicuity, and efficiency, also providing better dependability and stimulation.} \label{ueq_analysis}
          Two models can be compared relatively easily through a statistical comparison of UEQ answers \cite{ueq}. In this study, only 15 relative items were selected from 26 items to evaluate the quality of interactive software models; for more details (see Fig. \ref{fig:UEQ}) and \ref{experiment questionnaires}.\ref{ueq_sec}. We found that the 2D AGM was significantly higher than PTM in terms of attractiveness (p = 0.0026), efficiency (p = 0.036), and stimulation  (p =0.035). However, we found that there are no significant differences between the two models in terms of perspicuity (p=0.059) or dependability (p=0.113), see Table \ref{tab:ueq_table}. This point relates to \hyperref[RQ4]{RQ4} and \hyperref[RQ5]{RQ5}.
    
    \begin{figure*}[!ht]
          \begin{subfigure}{\linewidth}
           \centering
        \fbox{\includegraphics[width=0.87\textwidth]{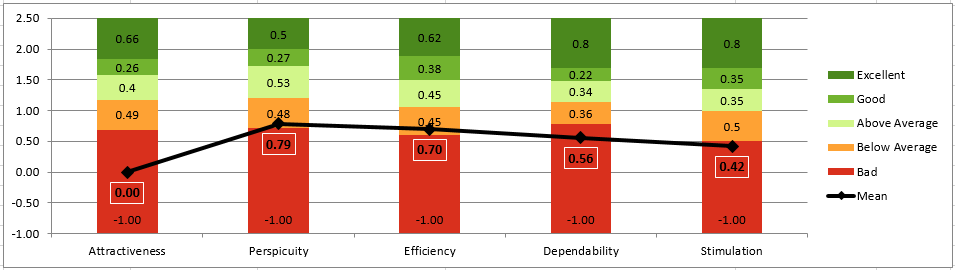}}
              \caption{The Benchmark of PTM Model}
              \label{fig:benchmark_PTM}
          \end{subfigure}
         \begin{subfigure}{\linewidth}
          \centering
              \fbox{\includegraphics[width=0.87\textwidth]{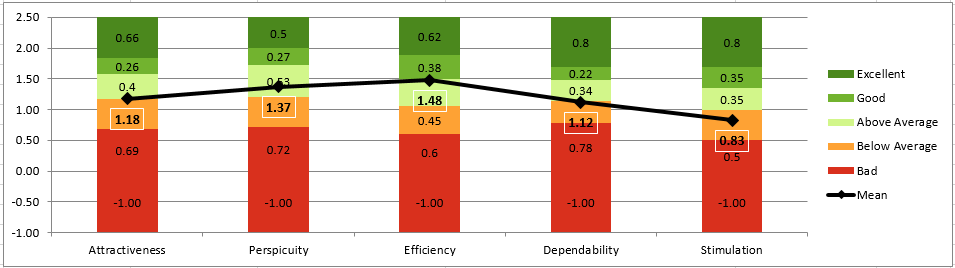}}
              \caption{The Benchmark of 2D AGM Model}
              \label{fig:benchmark_AGM}
          \end{subfigure}
             \caption{The UEQ Benchmark for Two Models. Differences are measured using the (t-test) mean values from the UEQ analysis tool. 2D AGM has below-average results in attractiveness (1.18), dependability (1.12), and stimulation (0.83); and above-average in perspicuity (1.37), and efficiency (1.48). The colours show the benchmark ranking of the mean values, and the black dots show the mean values achieved by the models on each scale.}
      
        \label{fig:benchmark}
        \end{figure*}
        
            \subsection{The User Experience Questionnaire UEQ (Comparing with UEQ benchmark dataset)\large{The 2D AGM provides an average level and PTM provides a below-average level compared to the UEQ benchmark.}} \label{benchmark_analysis}
           Figures \ref{fig:benchmark} show the comparison of the results for the two evaluated models against the benchmark. This allows conclusions to be drawn about their relative quality, and they can be compared to other similar models. With regard to the benchmark in terms of mean and median values (from -3 to 3), the UEQ analysis tool (using a t-test) showed that PTM has below-average results in perspicuity (0.79, 1) and efficiency (0.70, 0.5), and it has poor results in attractiveness (0.00, -0.5), dependability (0.56, 0.0), and stimulation (0.42, 0.5). On the other hand, 2D AGM has below-average results in attractiveness (1.18, 1), dependability (1.12, 1.5), and stimulation (0.83, 0.75), but it is above average in perspicuity (1.37, 1.33), and efficiency (1.48, 1.5), see Table \ref{tab:ueq_table} and Table \ref{tab:allResults}. Consequently, overall, 2D AGM provides an average level of user experience compared to the UEQ benchmark dataset, related to \hyperref[RQ4]{RQ4} and \hyperref[RQ5]{RQ5}. 
    
        \subsection{The NASA-TLX Questionnaire (Measuring Workload): The 2D AGM involves a significantly lower workload than the PTM.}\label{nasa_tlx_analysis}
        
        The NASA-TLX is an effective technique for assessing relative workload levels and provides subjective ratings of workload. The workload can be defined as the difference between the ability of workers and the demands of the job and it may be caused by many different factors. We evaluate several of them individually by using NASA Task Load TLX\cite{nasa}, see Appendix \ref{experiment questionnaires}.\ref{nasa tlx sec}.  \\
        We observed that 2D AGM requires a significantly lower overall workload than PTM (48.33 vs 64.33, Wilcoxon signed-ranks test: p = 0.01*; see Table \ref{tab:allResults}, Fig. \ref{fig:workload_plot}) which is related to \hyperref[RQ4]{RQ4} and \hyperref[RQ5]{RQ5}.
        \newline

         \subsection{ Completion Time Differences: PTM requires significantly less time to interpret an argument than 2D AGM}\label{time_analysis}
\begin{figure*}[!ht]
                \centering
                \begin{subfigure}{.5\linewidth}
                  \centering
                 \fbox{\includegraphics[width=0.95\textwidth] {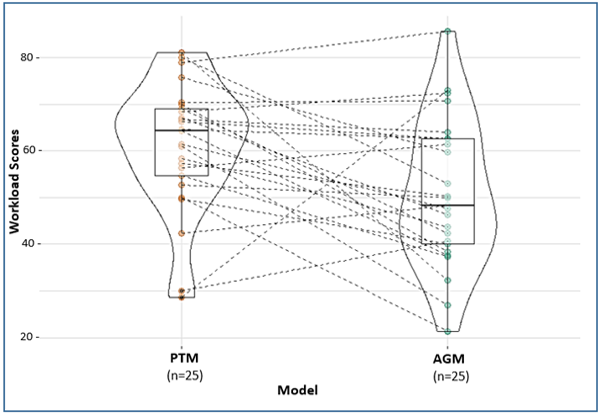}}
                     \caption{The paired plot of NASA TLX overall workload scores for two models.
                     }
                     \label{fig:workload_plot}
                \end{subfigure}%
                \begin{subfigure}{.5\linewidth}
                  \centering
                    \fbox{\includegraphics[width=0.95\textwidth] {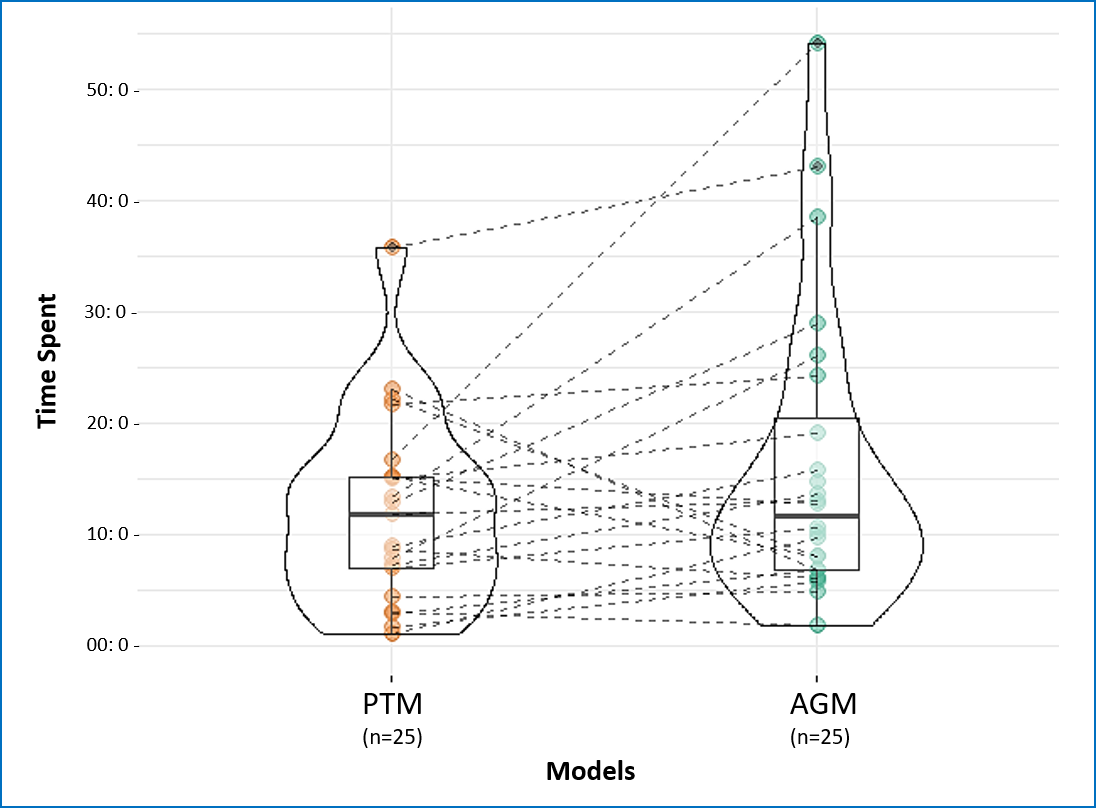}}
                     \caption{The paired plot of time spent on the two models. 
                     }
                   
                     \label{fig:spentTime}
                \end{subfigure}
                \caption{The two paired plots (a) a plot of NASA TLX overall workload for two models. A significantly lower workload score (p = 0.01) was achieved by participants in the 2D AGM. (b) a plot of time spent on two models. A significantly lower time spent (p = 0.015) was achieved by participants in the PTM.  Dashed black lines connect scores for individual participants.
                }
                
                \label{fig:nasa_time}
         \end{figure*}

         We found that the PTM requires significantly less time to interpret an argument than 2D AGM (median 11:13 min vs 13:05 min; Wilcoxon signed-rank test: p=0.015; see Fig. \ref{fig:spentTime} and Table \ref{tab:allResults}). This relates to \hyperref[RQ4]{RQ4}. However, 2D AGM introduces a newly proposed design and configuration (a new mechanism for interpreting an argument). Therefore, increasing interaction and addressing the familiarity gap with the model could reduce this performance gap. This result is consistent with \cite{Riechert18}, who found that model performance is influenced by familiarity and practical use. 

        \subsection{ Time spent and data size/argument length: Time spent is proportional to argument length} \label{time and data size}
\begin{figure*}[!ht]
                \centering
                \begin{subfigure}{.5\linewidth}
                  \centering
                  \fbox{\includegraphics[width=0.95\textwidth]{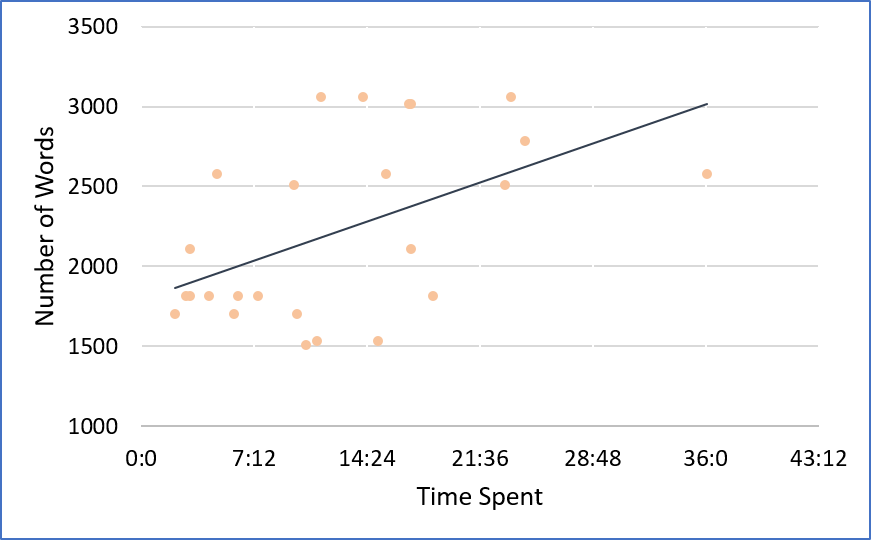}}
                  \caption{The PTM Model}
                  \label{fig:linear_ptm_td}
                \end{subfigure}%
                \begin{subfigure}{.5\linewidth}
                  \centering
                  \fbox{\includegraphics[width=0.95\textwidth]{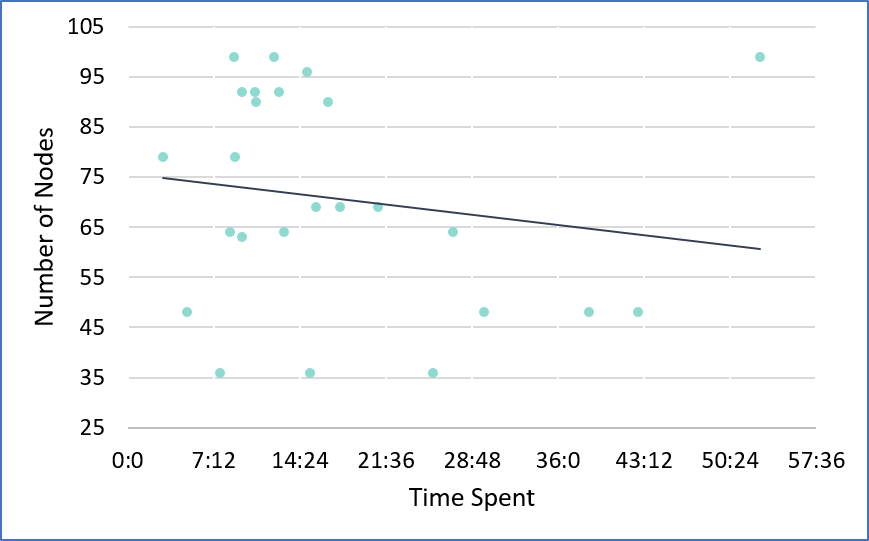}}
                  \caption{The 2D AGM Model}
                  \label{fig:linear_AGM_td}
                \end{subfigure}
                \caption{The linear regression of time and data size/argument length on both models. There is a correlation in PTM (p=0.01*) between time and the number of words in the argument, but no correlation in 2D AGM (p=0.44).
                }
               
                \label{fig:linear1}
         \end{figure*}

        We use linear regression between the time spent on PTM and 2D AGM and the length of the AG representation, which will be the number of words in PTM and the number of nodes in 2D AGM (see Fig. \ref{fig:linear1}). We observed that time spent on PTM is proportional to the number of words in the argument (p=0.01, $R^2 =0.258 $), see Fig. \ref{fig:linear_ptm_td}. On the other hand, we observed that no relation between time spent on 2D AGM and the number of nodes in the graph representation (p=0.44, $R^2 =0.0263$), see Fig. \ref{fig:linear_AGM_td}. This is related to \hyperref[RQ3]{RQ3} and \hyperref[RQ4]{RQ4}. 
    
         \subsection{Time spent and workload: No correlation between workload and time spent on both models} \label{time_workload}
\begin{figure*}[!ht]
                \centering
                \begin{subfigure}{.5\linewidth}
                  \centering
                  \fbox{\includegraphics[width=0.95\textwidth]{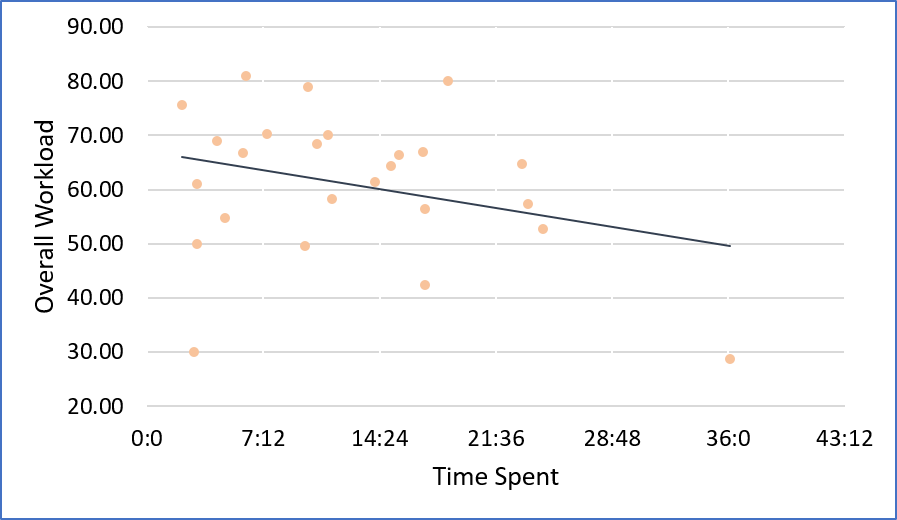}}
                  \caption{The PTM Model }
                  \label{fig:linear_PTM_tw}
                \end{subfigure}%
                \begin{subfigure}{.5\linewidth}
                  \centering
                  \fbox{\includegraphics[width=0.95\textwidth]{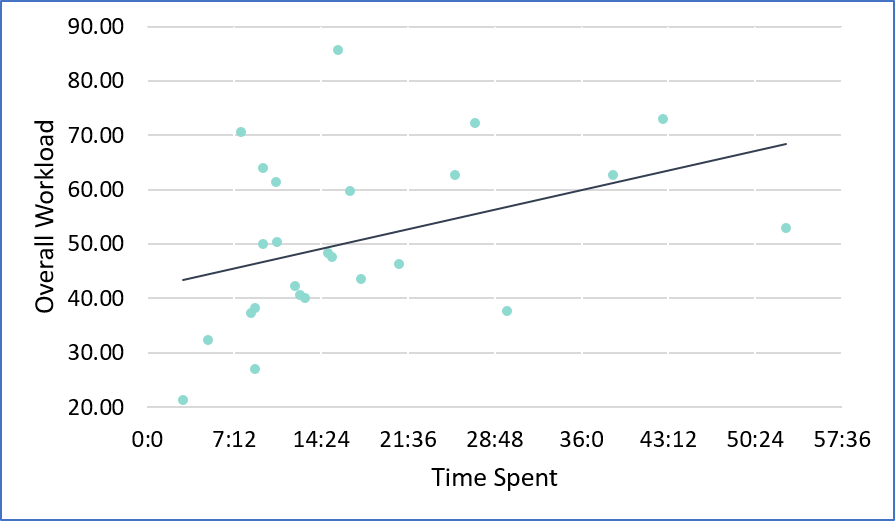}}
                  \caption{The 2D AGM Model }
                  \label{fig:linear_AGM_tw}
                \end{subfigure}
                \caption{The linear regression of the time spent and the workload on the two models. There is no significant correlation between the workload and the time spent on PTM (p=0.15) or 2D AGM (p=0.05).
                }
               
                \label{fig:linear2}
        \end{figure*}

         We used linear regression between the time spent on PTM and 2D AGM and the overall workload. We found that there is no significant correlation between the workload and the time spent on PTM (p=0.15, $R^2 = 0.089$) or 2D AGM (p=0.05, $R^2 = 0.157$), see Fig. \ref{fig:linear2}. This is related to \hyperref[RQ3]{RQ3}, \hyperref[RQ4]{RQ4}, and \hyperref[RQ5]{RQ5}.

    \subsection{ Feedback on the experiment:} \textbf{The participants found the experiment helpful\footnote{It helps and supports the interpretation process and shows various arguments from two different perspectives.} and valuable \footnote{It has pieces of important, untried information and methods that are interesting to know.}} \label{feedback} \\
     Table \ref{tab:feedback1} shows the overall feedback on the experiment from the participants. The participants found the experiment was considered to be valuable (N=23, 92\%),  helpful (N=23, 92\%), and liked (N=21, 84\%) by the participants, assessed under a Likert-scale value (>4). These results relate to \hyperref[RQ1]{RQ1} - \hyperref[RQ6]{RQ6} . See  \ref{feedback_comments}, which contains a sample of feedback comments.

              \begin{table} [!ht]
              \centering
               \caption{The feedback from 25 participants after completing the experiment. Answers on Likert Scale [1-7], 1 - strongly disagree, 7 - strongly agree}
               \label{tab:feedback1}
                \begin{tabular}{p{5.5cm} | p{1.5 cm} }
                \toprule
                   & \textbf{Feedback},  Median [Q1,Q3]\\
                  \midrule
                  I liked the Experiment & 6.0 [5, 7]\\
                  The experiment is helpful  & 6.0 [5, 6]\\
                  The experiment investigates a valuable topic  & 6.0 [5, 6] \\
              \bottomrule  
              \end{tabular}
          \end{table}
\section{Discussion}
\subsection{Overall results discussion and synthesis}
We set out to understand the impact of enriched argument representations on the process of critically interpreting argumentation and debates. As a representation paradigm, we focused on relations and categories that are prevalent in argument representation, whose extraction is being `commodities' with the recent evolution of argument mining methods. As expected, the introduction of a new representation modality that competes with a current text-based representation implies an initial learning and additional interpretation overhead to the end-users. Differences in time spent by the participants on each model are considered temporal measures reflected in this overhead ( p = 0.015, \ref{time_analysis}). In contrast, the structured modality (2D AGM) significantly outperformed the PTM in terms of attractiveness,  stimulation and efficiency, also providing better dependability and perspicuity (RQ5 - What are the users' perceptions of their preference for, and the easiness, effectiveness, and usefulness of the proposed solution?). Additionally, the 2D AGM significantly outperformed the PTM with regard to baseline user experience benchmarks, with a particular emphasis on perspicuity and efficiency. These results provide evidence that a visualisation model which elicits the structure of arguments can support the interpretation of complex arguments and debates (addressing RQ1 – Can discourse-level graph representations support the interpretation of complex arguments and debates? and RQ3 – Can argumentation schemes and argumentation types support the construction of this representation?).

We also confirmed the inherent trade-off between the time required to interpret an argument and critical interpretation, hypothesising that the additional structure, while introducing a temporal overhead, forces users to be more critical and systematic about the quality of the argument (claim-premise structure, argumentation schemes and supporting CQs).
Aiming to answer RQ4 (How does the proposed model perform for recall and workload metrics when compared to a pure text-based interpretation?), we found that the 2D AGM also required significantly less workload from the users than the PTM (p = 0.01, \ref{nasa_tlx_analysis}). Additionally, there was some level of evidence (although not statistically significant) that the graph-based representation may scale better for longer arguments when compared to PTM (p = 0.01, \ref{time and data size}). Both models were found to have similar effects in relation to working memory, as measured by a recall task (p = 0.247, \ref{BLEU_analysis}).

Being able to be critical about an argument implies a direct dialogue with the ability to ask CQs about the parts of an argument. We also controlled for the relationship between the number of CQs instantiated in the context of an argument and potential biases and initial positions, such as the level of perceived familiarity with the topic, and level of controversy. The higher the perceived familiarity, the lower the number of CQs, although this does not explicitly affect the interpretation process (addressing RQ2 – Which categories and relationships positively affect the interpretation process?).

Our findings suggest that the model configuration plays different roles in influencing participants’ interpretation of the argument, especially with regard to change of position on the topic, and that it has more influence on unfamiliar argument topics. Thus, we begin by discussing the model’s design implications and limitations, the results that were similar for models, and the slight impact that the 2D AGM model had compared with PTM. Finally, we discuss the positive impacts of 2D AGM and the benefits of the experiment settings.

\subsection{2D AGM design implications and limitations}
In our findings, we saw that the model’s performance (time spent) is observed in practice by the participants’ understanding and, the complexity of the argument topic, yet it is hardly affected by the model’s design. This indicates the potential need for some support for participants so that they can better understand the design and its related components (node, scheme, and relation types). Then, the key message that needs to be conveyed to participants takes less time to read a more complex topic. For instance, on the least complex topic, the participant who spent the shortest length of time on the task had not spent time reading all the premises because a pre-agreement evaluation was made. On the other hand, on the most complex topic, the slowest participant spent a long time reading all the premises, even though the pre-agreement evaluation had already been made, see Supp. Fig. \ref{fig:allspentTime}. Obviously, a significantly different amount of time was spent on two different complex topics. In other words, it is critical for participants to estimate their ability to read the argument and understand the model design in order for the model to have an honest evaluation.
To be realistic, there were other reasons for the long time spent on 2D AGM. These results are a normal and expected reaction to initial interaction with an unfamiliar graph, based on Walton’s classification schemes. The graph has some of the particular specifications of the representation’s design and new configurations. The two levels of difficulty caused by the model’s unfamiliarity and its configuration led to an increase in temporal and physical demand, see  \ref{time_analysis}. Fortunately, neither the workload nor the data were significantly affected by time, see \ref{time and data size} and \ref{time_workload}. 

\subsection{Similar results to those of PTM}
According to various metrics in our experiment, such as the A-BLEU score, we also discover that the effects of the 2D AGM model are comparable to those of the PTM. Working memory is one of the key findings, and the A-BLEU score is used to measure it (see Result 6.3). It is difficult to see significant differences between the two models because the A-BLEU is based on the similarities between the candidates (recall sentences) and references, according to the ability of the participants to recall the sentences, see \ref{tab:reclaims_sentences}. Although the 2D AGM model configurations help to break down the text into sentences, categorise them into different types, connect them as a relational tree, and attach premises with scheme types, the recall sentences are based on the participant's ability to recall and are not influenced by the model design. For instance, the BLEU score on 2D AGM is 0.05 for the most complex topic ``Abortion'', which is quite lower than PTM (0.12). On the other topic low in complexity, ``Tablets and textbooks'', the 2D AGM model (0.2) is much higher than PTM (0.02). On the non-complex or simple topic, ``Zoos'', 2D AGM (0.09) is quite similar to PTM (0.095), see \ref{fig:bleu-results-all}. 

\subsection{Slight impact of 2D AGM}
In other findings, the participants were asked for their opinion before and after reading the topic in each model. This evaluation helps us to know whether the model has an effect on the participants’ decisions or not. The fact that 2D AGM represents arguments in a different way than the PTM, as the participants are able to see where the premises came from and how their claims undercut or rebut other claims, for example. The participants can increase their confidence in the premises by asking the attached CQs or similar questions that test the acceptability of the premises’ statements. Moreover, 2D AGM shows which side of the argument contains more supporting nodes. Furthermore, the agreement evaluation is based on the main individual claims on each side and then the total points show the final decision, which is closer to the participant’s decision than a one-time decision. The decision is made based on the accumulative process of the partial parts of an argument on 2D AGM, rather than a one-time decision made on PTM, see (\ref{change decision}). This makes the decision a more reasonable, fair decision and close to the nature of the debate process. Combined, these reasons mean that 2D AGM has more alternative decisions than PTM. Actually, at this point, we can still see a slight impact of the 2D AGM design over the PTM.

\subsection{The performance of AGM compared with PTM}
Our findings in terms of user experience, workload, and user perception (preference) showed that the 2D AGM outperformed PTM, with significant differences between the two models, see  \ref{opinion_analysis}, \ref{ueq_analysis}, \ref{benchmark_analysis}, and \ref{nasa_tlx_analysis}. These findings may contain bias that mainly influences the measurements but is unlikely to be a confounding variable for our main results regarding whether or not the model affects the user experience or workload. This is because all of our analyses were only conducted with two standard questionnaires, which are based on several quantitative items and scales, see \ref{fig:benchmark} and \ref{tab:ueq_table}. Additionally, our experimental tasks, agreement evaluation, and other queries had been designed in the same way without any bias towards either model. Therefore, we believe that our results have a high level of reality and reliability, see \ref{fig:nasa-tlx} and \ref{fig:UEQ}. 

\subsection{Only on 2D AGM: Number of selected CQs correlated to the topic's familiarity}
The most essential point is the decision made by the reader and whether or not it is based on confidence and acceptability reasons. Thus, we observe that the argument or debate can be evaluated using other measurements that are the natural assessment criteria for argument and debate, such as familiarity and the argument’s representation. For instance, if you have two arguments with different levels of familiarity, then the reader will be more confident and assured about the decision made about the more familiar argument topic. Similarly, the way an argument is read and the supporting items or tools can positively or negatively affect the decision on that argument. For example, they can help to make the reader more confident, by knowing the source of the premise statements or the majority of the reasons on each side. In this study, we can see the impact of the representation of the argument. It is one of the interesting results that the method of representation can make the reader more confident, even on unfamiliar topics, so the reader does not need to ask many CQs, see \ref{CQs_analysis}.

\subsection{Benefits of using the ``Uni-Excel'' plugin package} \label{Uni-Excel}
The project was shown to the participants on a desktop and they shared access using Microsoft Teams; they had to have a {anonymous} account. The project embedded the “Uni-Excel” plugin package, which has excellent benefits and is useful for collecting, organising, and saving results during the experiment. Its main benefit is that if a technical issue occurs and interrupts the experiment, the participants do not need to start everything over again, just pick up where they left off, because the results are collected after each frame and this saves time. This happened to three or four participants, either due to internet connection issues or other unexpected disconnecting issues. 

\section{Future Work}
With the empirical evidence obtained in this study, we would like to conduct more confirmatory studies in the future to validate the results and examine their influence in different settings and representations, for different populations, and for different tasks. Indeed, we will investigate, with more rigorous argument representation, how the types of argument schemes, which will be explicitly represented, will affect the critical interpretation process. The study will base its findings on the subtle, scientific, analytical, and spontaneity measurements for human interaction. Ultimately, we acknowledge that interpretation is a complex concept and multidimensional construct. Though we attempted to measure interpretation using a set of metrics including the NLP functions, rational tasks, and survey methods, future studies are needed to understand how to critically interpret a complex argument or debate.

\section{Conclusions}
As argumentation becomes more complex, forked, and expanded in everyday life, investigating how humans interpret an argument critically becomes increasingly important and needs effective ways of interpretation other than textual. This study investigated whether a structured representation of an argument (2D AGM), can improve interpretation in the context of argumentative discourse when contrasted with arguments in textual form. The study included an experiment that presented 10 argument topics in two models: PTM and a 2D AGM model. The evaluation process for these models was based on task-related constructs (time spent, working memory, effort, and workload) and user experience (twenty-five participants). The structured modality (2D AGM) significantly outperformed the text-based one (PTM) in terms of attractiveness, perspicuity, and efficiency, also providing better dependability and stimulation. These results provided evidence that a visualisation model, which elicits the structure of arguments, can support the interpretation of complex arguments and debates. We also confirmed the inherent trade-off between time spent interpreting an argument (higher for 2D AGM when compared to a textual modality) and critical interpretation, hypothesising that the additional structure, while introducing a temporal overhead, forces users to be more critical and systematic when judging the quality of the argument. We also found that the 2D AGM implied significantly less workload for users when compared to the PTM. Additionally, there was some level of evidence that the structured model may scale better for longer arguments than the PTM. Both models were found to have similar effects in relation to working memory. Future work could involve gaining a better understanding of which categories within an argument scheme are most relevant for critical interpretation.

\section*{Acknowledgements}
 We thank all the volunteers and participants, and all supporting staff, who provided helpful comments on this study and experiment. The authors gratefully acknowledge all people who worked with them.



\printbibliography 
\clearpage

\appendix
\setcounter{secnumdepth}{0}
\setcounter{table}{0}
\renewcommand\thetable{A.\arabic{table}}
\setcounter{figure}{0}
\renewcommand\thefigure{A.\arabic{figure}}

\section{APPENDICES}
In this section, we present additional background and details about the argument models.
\subsection{Argumentation Structure: Random Topics}   \label{argument topics}
  Different debate topics are collected from the `procon.org' website. Then, selected 10 different topics which are listed in the different areas as the following:  
\begin{enumerate}
    \item \textbf{Education:} 
     Banned books (Parents or other adults should be able to ban books from schools and libraries.) and  Tablets vs Textbooks (Tablets should replace Textbooks in High Schools.)
    \item \textbf{Health and Medicine:}
    Obesity (Obesity is a disease.), Abortion (Abortion should be Legal.) and Vegetarianism (People should become vegetarian).
    \item \textbf{Science and Technology:}
    Battled water (Bottled water should be banned.), Animal testing (Animals should be used for scientific or commercial testing.), Climate change (Human activity is primarily responsible for global climate change.), Zoos(Zoos protect animals. Zoos should exist.), and Energy (Alternative energy effectively replaces fossil fuels.)
\end{enumerate}
\subsection{Argument Topics Complexity}
Using python code to calculate the complexity of the arguments. We measure the total number of words, the total number of sentences, the average word frequency, and the word average length. These are some basic metrics that can be used to quantify the characteristics of a text and can provide insights into the language and content of the text, see Figure \ref{fig:argument_complexity_count}. 
We found that the complexity of the arguments can be measured in terms of lexical diversity\footnote{"Lexical diversity" is a measure of the variety of words used in a text, calculated as the number of unique words divided by the total number of words.}, lexical density\footnote{ "Lexical density" is a measure of the proportion of content words to function words in a text. Content words are words that carry the meaning of a sentence, while function words serve to connect content words together.}, dis-similarity\footnote{ "Dis-similarity" is a measure of the difference between two texts, calculated as the proportion of unique words in one text that does not appear in the other.} and connection words ratio\footnote{"Connection words ratio" refers to the proportion of words that connect ideas or sentences in a text, such as conjunctions (e.g. "and", "but"), prepositions (e.g. "in", "on"), and pronouns (e.g. "he", "she").}. These metrics help to create coherence and cohesion within a text., see Figure \ref{fig:argument_complexity}.  
\newline
In terms of readability, we measure the argument topics in terms of the Flesch Reading Ease Score\footnote{The Flesch Reading Ease Score is a formula that computes the readability of a text based on its average sentence length and the average number of syllables per word.}, Spache Readability Score\footnote{The Spache Readability Score is a measure of the simplicity of a text based on sentence length and the number of familiar words used. It is calculated using a formula that takes into account the number of syllables in a sentence and the number of easy words used. The score ranges from 0 to 100, with a higher score indicating greater readability.}, Dale-Chall Score\footnote{The Dale-Chall readability Score is a readability metric that measures the readability of a text by taking into account the frequency of unfamiliar words in a text and the length of the words used. A Dale-Chall score of 0 would indicate that a text is very easy to read and contains very few or no unfamiliar words.} and SMOG Score\footnote{The Simple Measure of Gobbledygook (SMOG) is another formula that estimates the readability of a text by counting the number of polysyllabic words and applying a mathematical formula. SMOG is a popular method to use on health literacy materials.}. These are four different readability metrics used to evaluate the readability of text, see Figure \ref{fig:argument_complexity_read}. 

\subsection{Argumentation Structure: Walton's Critical Questions}  \label{Walton's CQs}
\begin{itemize}
    \item Argument from Expert Opinion 
    \begin{itemize}
        \item How credible is E as an expert source?
        \item Is E an expert in the A's field?
        \item What did E assert that implies A?
        \item Is E personally reliable as a source?
        \item Is A consistent with other experts assert?
        \item Is E's assertion based on cited evidence?
        \item Is A consistent with recent studies?
        \item Is A in the Conclusion's topic domain?
    \end{itemize}
    \item Argument from a Position to Know 
    \begin{itemize}
        \item Is ``P'' in position to know whether A is true or not?
        \item Is ``P'' an honest (trustworthy, reliable) source?
        \item Did ``P'' assertion implies that A is true or false?
        \item Is P's assertion based on cited evidence? 
    \end{itemize}
    \item Argument from Analogy
    \begin{itemize}
        \item Are there differences between C1 and C2 that would tend to undermine the force of the similarity cited?
        \item Is A true/false in C1?
        \item Which kind of similarity between C1 and C2?
    \end{itemize}
    \item Argument from Cause To Effect
    \begin{itemize}
        \item How strong is the causal generalization?
        \item Is the evidence cited strong enough to warrant the causal generalization?
        \item Are there other reasons that could interfere with the production of the effect in this case?
    \end{itemize}
    \item Argument from Positive/Negative Consequences 
    \begin{itemize}
        \item How strong is the likelihood that cited consequences will (must,may) occur?
        \item what evidence supports the claim that the cited consequences will (must,may) occur?
        \item Is it sufficient to support the strength of claim adequately?
        \item Is there other opposite consequences that should be taken into account?
    \end{itemize}
    \item Argument from Falsification
    \begin{itemize}
        \item Is it the case that if hypothesis is true, then a proposition reporting an event is true?
        \item Has a proposition reporting an event been observed to be true(false)?
        \item Could there be some reason why a proposition reporting an event is true, other than its being because of the given hypothesis true?
    \end{itemize}
\end{itemize}

\subsection{Experiment Models Tasks}  \label{tasks_methods}
\begin{enumerate}
    \item \textbf{ Task 1: Recall claims (the BLEU scores)} \label{task1_bleu_def}
 A-BLEU Score (A Bilingual Evaluation Understudy Score), is a metric procedure for evaluating a reference sentence against a user-generated sentence (candidate) and it indicates how similar or closer the candidate's sentence is to the reference sentence.  The BLEU metric ranges from 0 to 1, it finds legitimate differences in word choice and word order between the references and candidates, where score values closer to 1 represent more similarities.Start with a familiar metric, \textbf{Precision}, which is defined as the following:
          \begin{equation}
             Precision = \frac{No. Of Candidate Words Occurring In Any Reference}{Total No. Of Words In The Candidates} 
          \end{equation}
The BLEU metric ranges from 0 to 1, it finds legitimate differences in word choice and word order between the references and candidates, where score values closer to 1 represent more similarities. \textbf{Formally:}
          \begin{equation}
              BLEU = BP \times \exp  ({\sum_{n=1}^{N} w_{n}* \log(p_{n})}) 
          \end{equation}
Where BP is the brevity penalty term, which will be 1 if the candidate length in the same as any reference length, N is the number of n-grams, $p_{n}$ is the modified precision, and $w_{n}$ is the corresponding weights, which are usually uniform in practice. \newline
In order to address the match between references and candidates, by comparing the n-gram precision of the candidates with the n-gram precision of the references, and using the form of modified precision to compare multiple candidates against multiple references.
         
\item \textbf{ Task 2: Selection CQs}  \label{task2_CQS}
 We computed the total number of selected critical questions, Formally: 
          \begin{equation}
              Total_Matches = {\sum_{n=1}^{N} COUNTIF (R_{1 \to m, C})} 
          \end{equation}
 Where $N$ is the total number of critical questions, $C$ is the criteria (the critical question was checked (answered) with ''TRUE"), and $R$ is the rows that contain critical questions. 
 \end{enumerate}
 
 \subsection{Experiment Questionnaires} \label{experiment questionnaires}
\begin{enumerate}
 \item \textbf{ NASA TLX Questionnaire} \label{nasa tlx sec}
The measurement of the NASA-TLX method is divided into two stages, namely the comparison of each scale (Paired Comparison) and assigning a value to the work (Event Scoring).
The NASA Task Load Index is a multi-dimensional rating procedure that provides an overall workload score based on a weighted average of ratings of six sub-scales: Mental Demands. Physical Demands, Temporal Demands, and Own Performance, Effort, and Frustration. This set of six rating scales was developed for evaluating a user's experiences during different tasks.  According to NASA-TLX\cite{nasa}, there are 15 possible source-of-workload comparison cards that contain pair-wise comparisons of six scales. The administering of the TLX involves two steps. \\ In the first step, a participant reflects on the task they’re being asked to perform and looks at each paired combination of the six scale factors to decide which is more related to their personal definition of workload as related to the task. This results in a user considering 15 paired comparisons. For example, they need to decide whether Physical Demand or Mental Demand ''represents the more important contributor to the workload for the specific task you recently performed” (see Figure \ref{fig:nasa-tlx1}). \\ The second step involves participants rating each of the six dimensions on scales from very low to very high. Each scale is presented by a slider that divides the scale from 0 to 100 in increments of 5 (see Figure \ref{fig:nasa-tlx2}). To calculate the workload of a task, each of the six scales is multiplied by the weight from step 1 to generate ''the adjusted rating," which is summed across six scales and divided by 15 to obtain the overall workload score for a participant in that task.
\item \textbf{ User Experience Questionnaire UEQ} \label{ueq_sec}
 The user experience questionnaire-UEQ\cite{ueq}, English version, consists of the scales of pragmatic and harmonic quality that are used to measure user experience and the quality of the interactive product. The UEQ can be used to compare the users' experiences of two products, usually two versions of one item: the old or established version, and the new version. It is used to determine whether the new version provides a better user experience. The items representing each scale were extracted from this dataset by principal component analysis. The items have the form of a semantic differential, i.e. each item is represented by two terms with opposite meanings. The UEQ uses a seven-stage scale to reduce the well-known central tendency bias for such types of items. The items are scaled from 1 to 7 where 1 represents the most negative answer, 4 is a neutral answer, and 7 is the most positive answer. The UEQ has 26 items with 6 different main scales, which are: attractiveness, perspicuity, efficiency, dependability, stimulation, and novelty(see Figure \ref{fig:UEQ}). 
\end{enumerate}

\newpage
\subsection{Supplementary Tables}        
  \begin{enumerate}
      \item \textbf{Table \ref{tab:reclaims_sentences} is for Task 1:Recall claims sentences.}
       \begin{table*}[h!]
          \caption{The examples of claim candidates for PTM and 2D AGM Task 1, which contain three candidate statements from the topic (Obesity) on the PTM model and four candidate statements from the topic (Tablet vs. Textbooks) on 2D AGM, got the highest BLEU scores. The table contains the participant project-generated IDs to show the individual results.}
          \label{tab:reclaims_sentences}
          \begin{tabular}{ll}
            \toprule
            A- ID T606 & PTM - Task 1: Recall claims (Candidate statements)\\
            \midrule
            1- & It is  a consequence of other factors, thus it is not a disease.\\
            2- & It is a disease ,  because it cause malfunctioning of other organs by  the productions of some  hormones.\\
            3- & It is not a disease,  because it can  be prevented by educating people on eating habits.\\
            \midrule
            B- ID T692 & 2D AGM - Task 1: Recall claims (Candidate statements)\\
            \midrule
            1- & Tablets introduce vision problems in students.\\
            2- & Tablets have are more environmentally friendly to produce.\\
            3- & Tablets cause greater distraction for students.\\
            4- & Studies show that students studying the same material on tablets attain higher test scores.\\
          \bottomrule
       \end{tabular}
    \end{table*} 
    
    \item \textbf{Table \ref{tab:PTM_claims} is for PTM Model (Task 1): Claims References statements.}
             \begin{table*}[h!]
          \caption{The example of claims references for PTM Task 1, which contains three references from the topic (Obesity) on PTM model for each side (Pro and Con) of the argument topic}
          \label{tab:PTM_claims}
          \begin{tabular}{ll}
            \toprule
             A- &  \textbf{PTM Claims Sentences in Pro/Agreement Side}\\
            \midrule
             1- & Reference1: It meets the definition of disease.\\
             2- & Reference2: It decreases life expectancy and impairs the normal functioning of the body.\\
             3- & Reference3: It can be caused by genetic factors.\\
            \midrule
             B- & \textbf{PTM Claims Sentences in Con/Disagreement Side}\\
            \midrule
             1- & Reference1: It is a preventable risk factor for other diseases.\\
             2- & Reference2: It is the result of eating too much.\\
             3- & Reference3: It is caused by exercising too little.\\
          \bottomrule
       \end{tabular}
    \end{table*}
    
\item \textbf{Table \ref{tab:AGM_claims} is for 2D AGM Model (Task 1): Claims References statements.}
      \begin{table*}[h!]
          \caption{The example of claims references for 2D AGM Task 1, which contains three references from the topic (Tablet vs. Textbooks) on the 2D AGM model on each side (Pro and Con) for the argument topic}
          \label{tab:AGM_claims}
          \begin{tabular}{p{1 cm} p{11 cm}}
            \toprule
             A- &  \textbf{2D AGM Claims Sentences in Pro/Agreement Side}\\
            \midrule
             1- &  Reference1: They are supported by most teachers and students, are much lighter than print textbooks, and improve standardized test scores.\\
            2- & Reference2: Tablets can hold hundreds of textbooks, save the environment by lowering the amount of printing.\\
            3- & Reference3: Increase student interactivity and creativity, and digital textbooks are cheaper than printed textbooks.\\
            \midrule
             B- & \textbf{2D AGM Claims Sentences in Con/Disagreement Side}\\
            \midrule
             1- & Reference1: They are expensive, too distracting for students, easy to break, and costly/time-consuming to fix. \\
            2- & Reference2: They say that tablets contribute to eyestrain, headaches, and blurred vision, which increases the excuses available for students not doing their homework.\\
            3- & Reference3: Require costly WiFi networks, and become quickly outdated as new technologies emerge.\\
          \bottomrule
       \end{tabular}
    \end{table*}
     
\item \textbf{Table \ref{tab:task2-result} is for 2D AGM Model (Task 2): Example of Selection CQs} \label{cqs-sec}
        \begin{table*}[h!]
            \caption{An example result of 2D AGM Task 2 contains 11 Critical Questions (CQs), which are picked up from Walton's Critical Questions (see Appendix \ref{Walton's CQs}). The table includes the CQs and selected options, where TRUE = selected CQs and FALSE = not selected CQs.}
            \label{tab:task2-result}
            \begin{tabular}{l l l}
                \toprule
              No & 2D AGM - Task 2: Recall Critical Questions CQs & Selected\\
                \midrule
                1- & How credible is Expert as an expert source?	& TRUE \\
                2- & Is Assert consistent with other experts assertion? &	FALSE \\
                3- & Is the PTK an honest (trustworthy, reliable) source? &	TRUE \\
                4- & Is PTK's assertion based on cited evidence? &	FALSE \\
                5- & Is it sufficient to support the strength of claim?  &	FALSE\\
                6- & Is there other opposite consequences?	& TRUE \\
                7- & How strong is the causal generalization? &	FALSE \\
                8- & Are there other reasons of the effect in this case?	& TRUE \\
                9- & Which kind of similarity between C1 and C2?	& TRUE \\
                10- & Is Assertion true/false in C1? &	FALSE \\
                11- & Is it the case that if hypothesis is true, then a proposition reporting an event is true? &	FALSE \\
              \bottomrule
           \end{tabular}
        \end{table*}
      
\item \textbf{Table \ref{tab:ueq_table} contains the UEQ analysis tool's results (by using a t-test) that compare between two models}
       \begin{table*}[h!]
              \caption{The User Experience Questionnaire (UEQ) results are for two models on five scales: attractiveness, perspicuity, efficiency, dependability, and stimulation. The table shows the scale means, standard division (STD), confidence, and corresponding  5\% Confidence Interval(CI) values of PTM and 2D AGM. The confidence interval is a measure of the precision of the estimation of the scale mean. The smaller the confidence interval is, the higher the precision of the estimation and the more trustworthy the result is. }
              \label{tab:ueq_table}
              \begin{tabular}{lllll}
               \toprule
               {\textbf{PTM Model}}\\
               Scales & Mean & STD & Confidence &  (5\%CI)\\
                   \midrule 
              Attractiveness & 0.00 & 1.63 & 0.64 & [-0.64,0.64] \\
              Perspicuity & 0.79 & 1.32 & 0.52 & [0.27,1.30] \\
              Efficiency & 0.70 & 1.54 & 0.60 & [0.10,1.30] \\
              Dependability & 0.56 & 1.44 & 0.56 & [0.00,1.12] \\
              Stimulation & 0.42 & 0.74 & 0.29 & [0.13,0.71]\\
              \\ \midrule
            \textbf{2D AGM Model}\\
                 Scales & Mean & STD & Confidence &  (5\%CI) \\
                    \midrule 
              Attractiveness & 1.18 & 0.82 & 0.32 & [0.86,1.50] \\
              Perspicuity & 1.37 & 0.73 & 0.29 & [1.09,1.66] \\
              Efficiency & 1.48 & 0.93 & 0.36 & [1.12,1.84]\\
              Dependability & 1.12 & 0.96 & 0.38 & [0.74,1.50]\\
              Stimulation & 0.83 & 0.59 & 0.23 & [0.60,1.06]\\
         \bottomrule
         \end{tabular}
        \end{table*}
   
   \item \textbf{ Table \ref{tab:feedback} contains all feedback comments by participants} \label{feedback_comments}
      \begin{table} [h!]
            \caption{Table contains two columns: the first column contains the project-generated id and the second one contains comments, which are provided by participants at the end of the experiment. }
            \label{tab:feedback}
            \begin{tabular}{p{1.5cm} p{6cm}}
                \toprule
                  \textbf{Participants ID} & \textbf{Comments} \\
                  \midrule
                  T756 & Thank you Hanadi for your interesting experiment and kind hospitality.. Good luck :)..\\
                  T298 & I like the idea of the model but GUI could be improved			\\
                  T606 & Thank you and good luck with your research :)			\\
                  T430 & I would recommend to future participants to use a mouse instead of a track-pad.		\\
                  T014 & I have never seen this style of argument be portrayed in this fashion. It was very helpful in making my own mind up about a controversial topic, Something that is often hard to using purely text-based arguments.		\\
                  T941 & frustration with mouse control and scrolling, difficult to explore everything, the layout of the nodes was a bit difficult and proximity made some things more prominent than others. Some sliders were confusing (e.g. which side is perfect). 						\\
                  T966 & AG graph looks very easy to read. 		\\
                  T062 & it was interesting but a bit long. \\
                  T809 & The instruction are often very unclear and the overall experience was a little frustrating. I often find that options were written in ways that do not help with understanding also.						\\
                  T115 & Maybe add arrows between the claims, to indicate what is being undercut.			\\
                  T020 & I was unfamiliar with the critical questions part and found it is difficult to understand.					\\
                  T498 & A better layout design may improve the user experience a lot.				\\
                  T177 & I really liked the second model in terms of clarity and organization, the only thing I would improve is the graphical visualization :)						\\
                  T375 & good \\
               \bottomrule
          \end{tabular}
      \end{table}
  \end{enumerate}

\clearpage
\newpage

\subsection{Supplementary Figures}       
\begin{enumerate}
\item \textbf{Figure \ref{fig:Arg-ex} is an example of the argumentation structure.}
\begin{figure*}[!ht]
    \centering
    \fbox{\includegraphics[width=0.9\textwidth, height=.50\textheight]{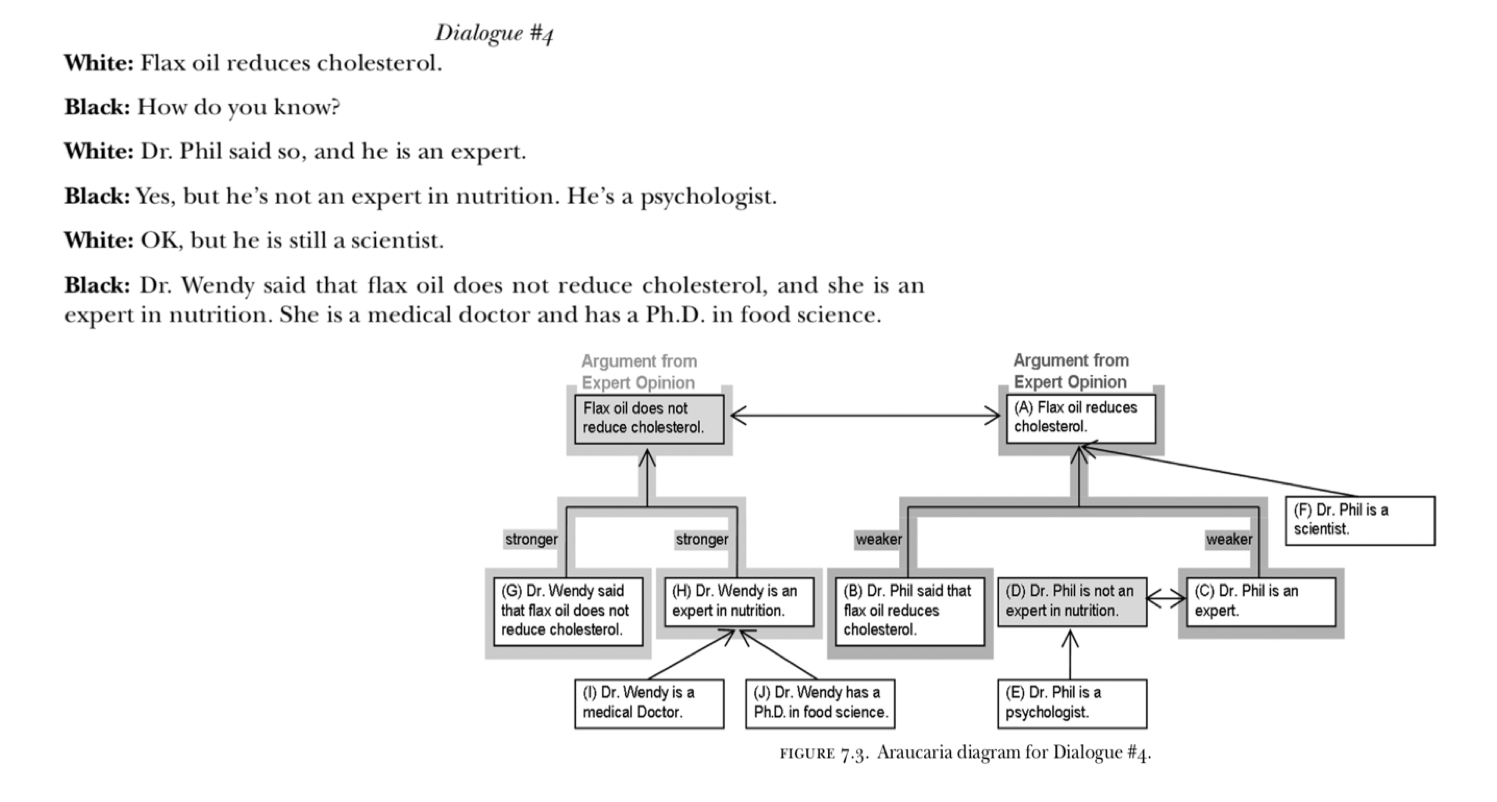}}
    \caption{An example of Walton's argumentation structure. It shows the argumentation structure that is about ``Flax oil reduces cholesterol".}
  
    \label{fig:Arg-ex}
\end{figure*}
\item \textbf{Figure \ref{fig:Arg-str} shows the argumentation structure by Milz\cite{Milz2017} which is based Walton's argumentation structure.} 
\begin{figure*}[!ht]
    \centering
    \fbox{\includegraphics[width=0.4\textwidth]{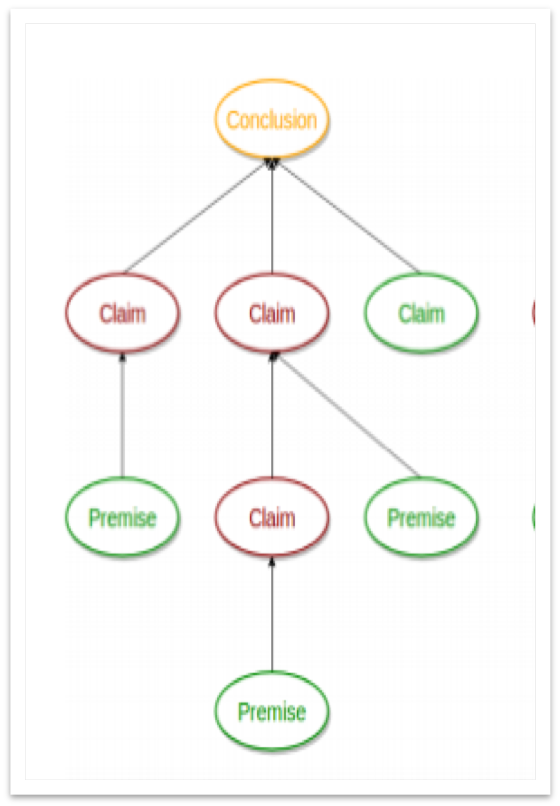}}
    \caption{A basic template example of Milz's\cite{Milz2017} argumentation structure, which contains the three main units in an argument: a conclusion, claim and premise node. }
    
    \label{fig:Arg-str}
\end{figure*}

    \item \textbf{Figures \ref{fig:argument_complexity_count}, \ref{fig:argument_complexity}, and \ref{fig:argument_complexity_read} respectively show the argument text complexity in terms of the total number of words, the total number of sentences, the average word frequency, and the word average length; in terms of lexical diversity, lexical density, pos-dissimilarity, and connection words ratio; in terms of Flesch Reading Ease Score, Spache Readability Score, Dale-Chall Score and SMOG Score. }
    \begin{figure*}[h!]
    \centering
    \fbox{\includegraphics[width=\textwidth]{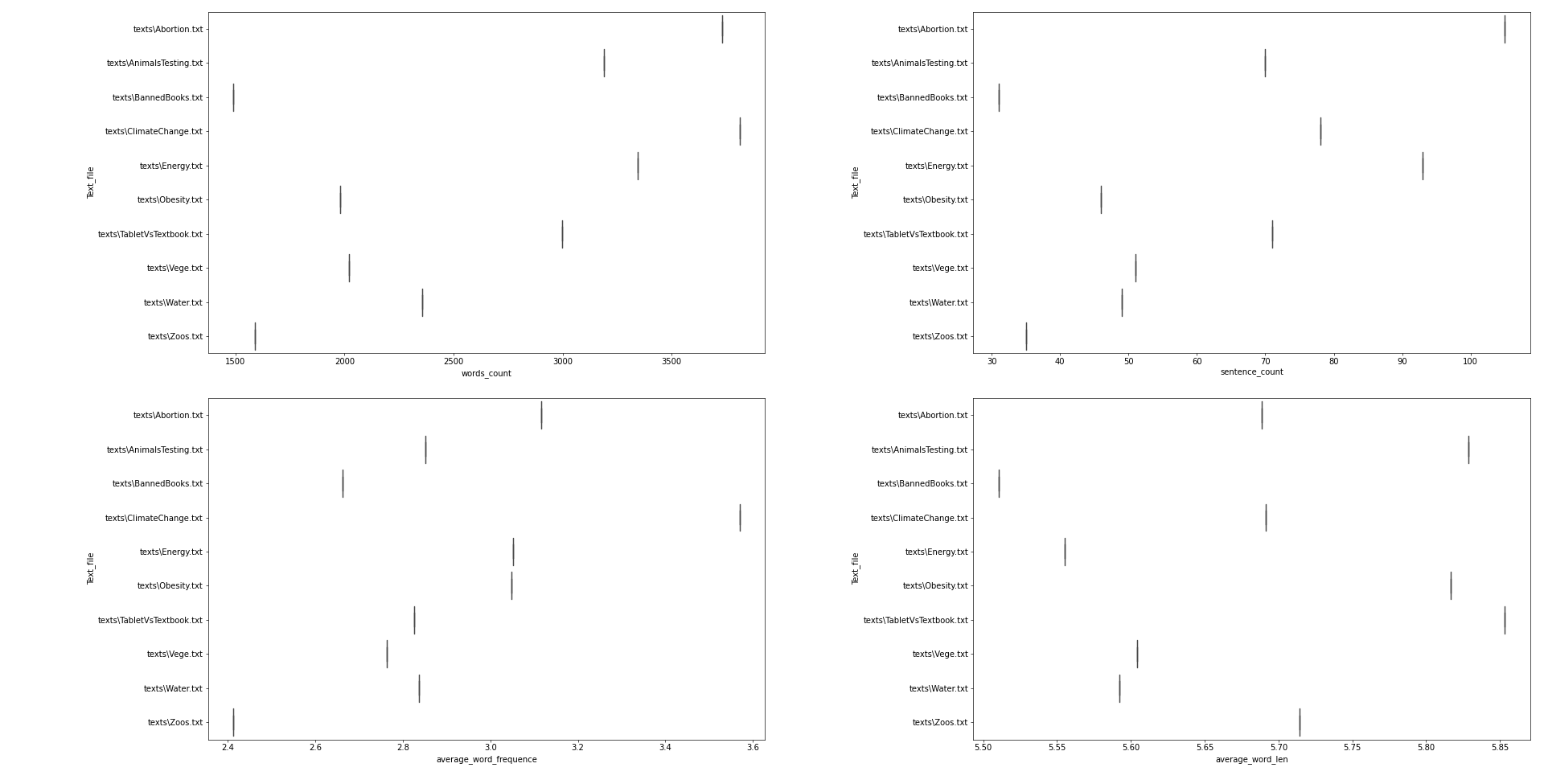}}
    \caption{Complexity of the 10 argument topics in terms of the total number of words, the total number of sentences, the average word frequency, and the word average length. }
    \label{fig:argument_complexity_count}
\end{figure*}
\begin{figure*}[!ht]
    \centering
    \fbox{\includegraphics[width=\textwidth]{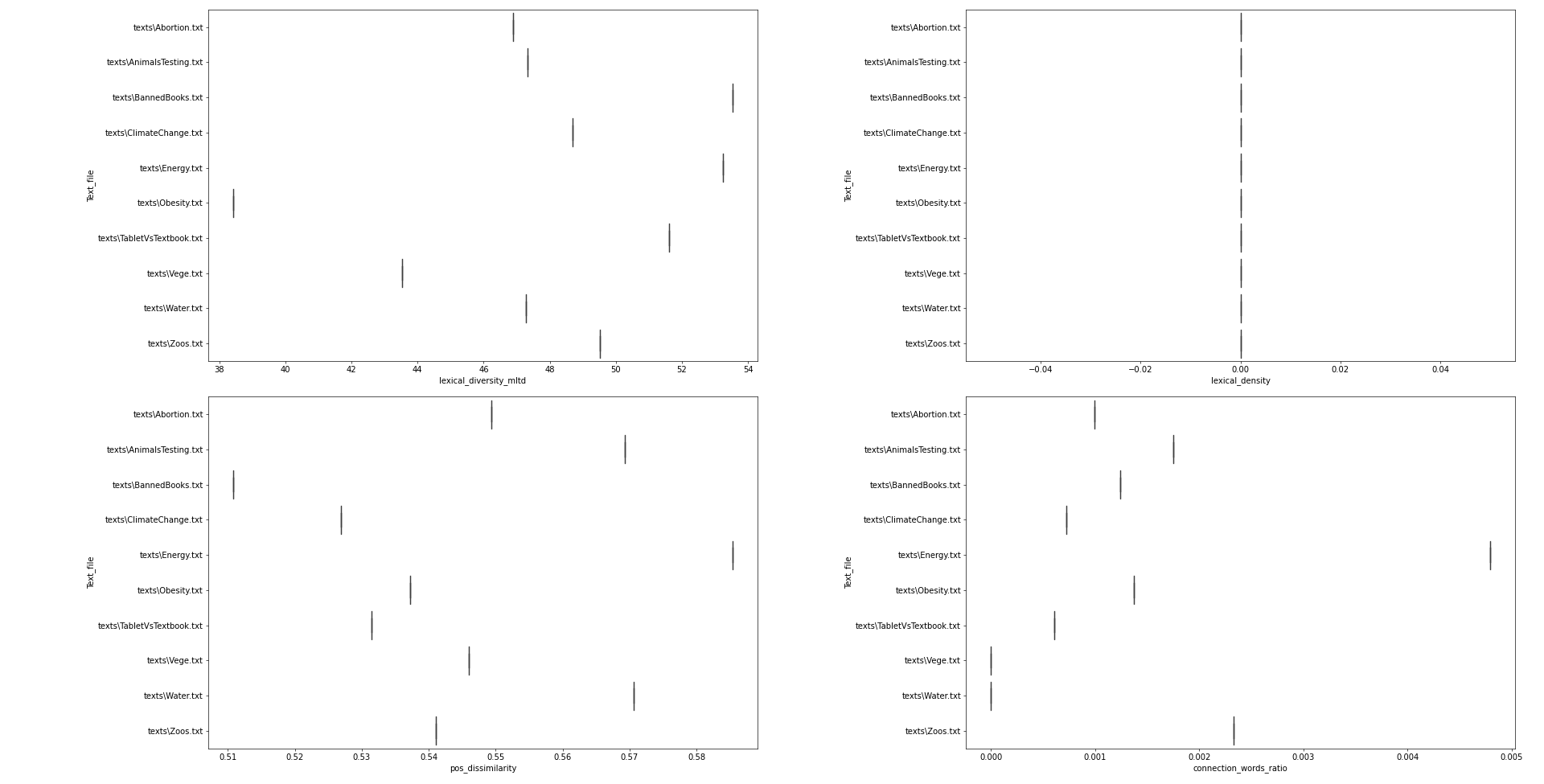}}
    \caption{Complexity of the 10 argument topics in terms of lexical diversity, lexical density, pos-dissimilarity, and connection words ratio }
    \label{fig:argument_complexity}
\end{figure*}
\begin{figure*}[h!]
    \centering
    \fbox{\includegraphics[width=\textwidth]{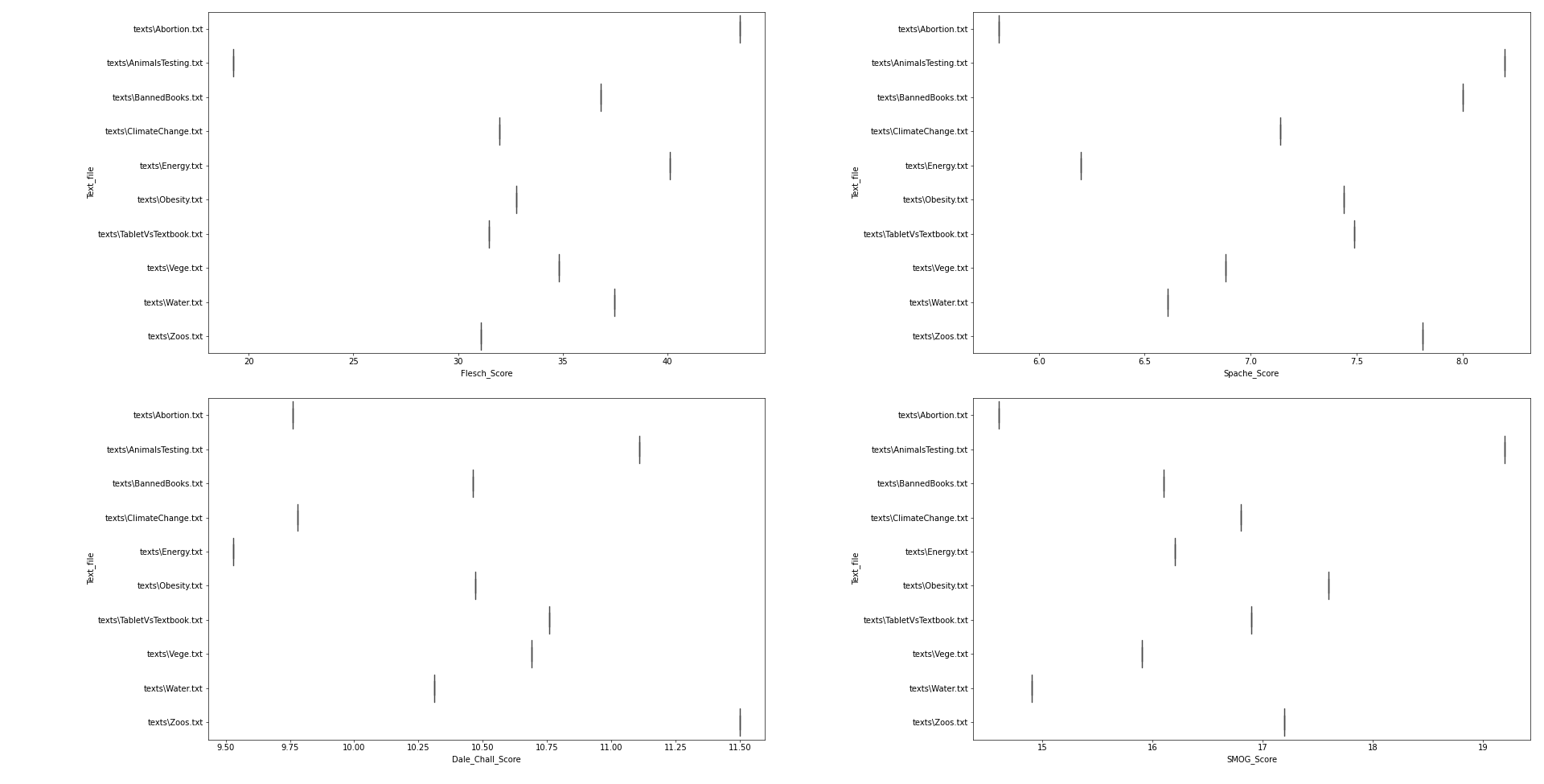}}
    \caption{Complexity of the 10 argument topics in terms of Flesch Reading Ease Score, Spache Readability Score, Dale-Chall Score and SMOG Score. }
    \label{fig:argument_complexity_read}
\end{figure*}
    \item \textbf{Figure \ref{fig:Evaluation-result} shows 2D AGM Model's evaluation sentences}
\begin{figure*}[!ht]
         \centering
         \fbox{\includegraphics[width=0.6\textwidth]{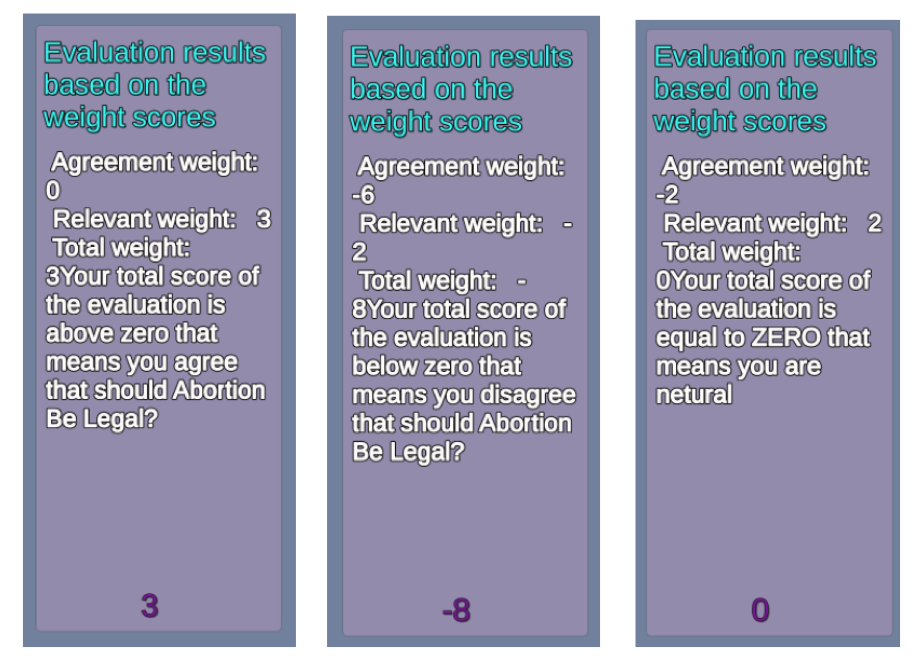}}
         \caption{The 2D AGM Evaluation result message, which shows the three different evaluation sentences, indicates the participant's decision, which is shown based on the aggregated score of (relevance and agreement) for main claims. For more details, see section \ref{agm score_aggregation}. From left to right: participant's decision agreed(3), disagreed(-8), and neutral with the topic.}
        
         \label{fig:Evaluation-result}
 \end{figure*}
 
 \item \textbf{Figure \ref{fig:bleu-results-all} shows a barchart for A-BLEU scores for all participants}
        \begin{figure*}[!ht]
             \centering
            \fbox{ \includegraphics[width=0.9\textwidth]{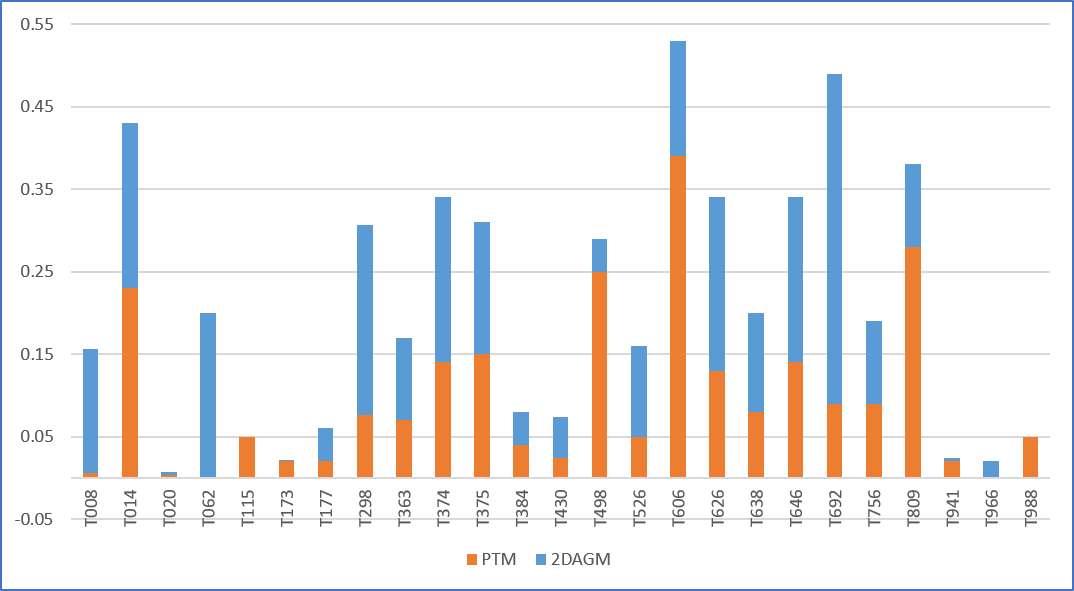}}
             \caption{The barchart shows the A-BLEU scores for all participants. The A-BLEU score (An Aggregate Bilingual Evaluation Understudy Score) indicates how similar the candidate's sentence is to the reference sentence. The A-BLEU metric ranges [0, 1], where score values closer to 1 represent more similarities.}
            
             \label{fig:bleu-results-all}
          \end{figure*}
          
\item \textbf{Figure \ref{fig:allspentTime} shows a barchart for the time spent by all participants }
                \begin{figure*}[!ht]
                     \centering
                    \fbox{ \includegraphics[width=0.9\textwidth]{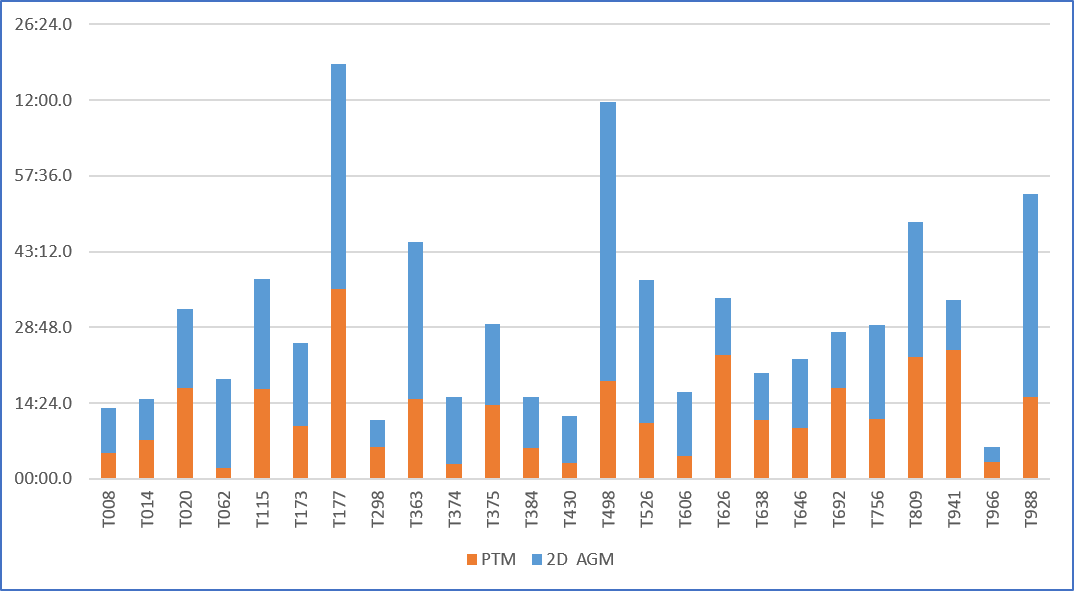}}
                     \caption{The barchart shows the participants' time spent in minutes and seconds. The maximum and minimum time are taken in PTM(36:04, 2:09) and 2D AGM(52:55, 2:56). The time spent is calculated by the hidden timer for each model using C\# code.}
                   
                     \label{fig:allspentTime}
                 \end{figure*}

\item \textbf{Figures \ref{fig:nasa-tlx} are for NASA TLX Questionnaire Panels}
  \begin{figure*}[!ht]
           \centering
             \begin{subfigure}{\linewidth}
               \centering
               \fbox{\includegraphics[width=0.9\textwidth]{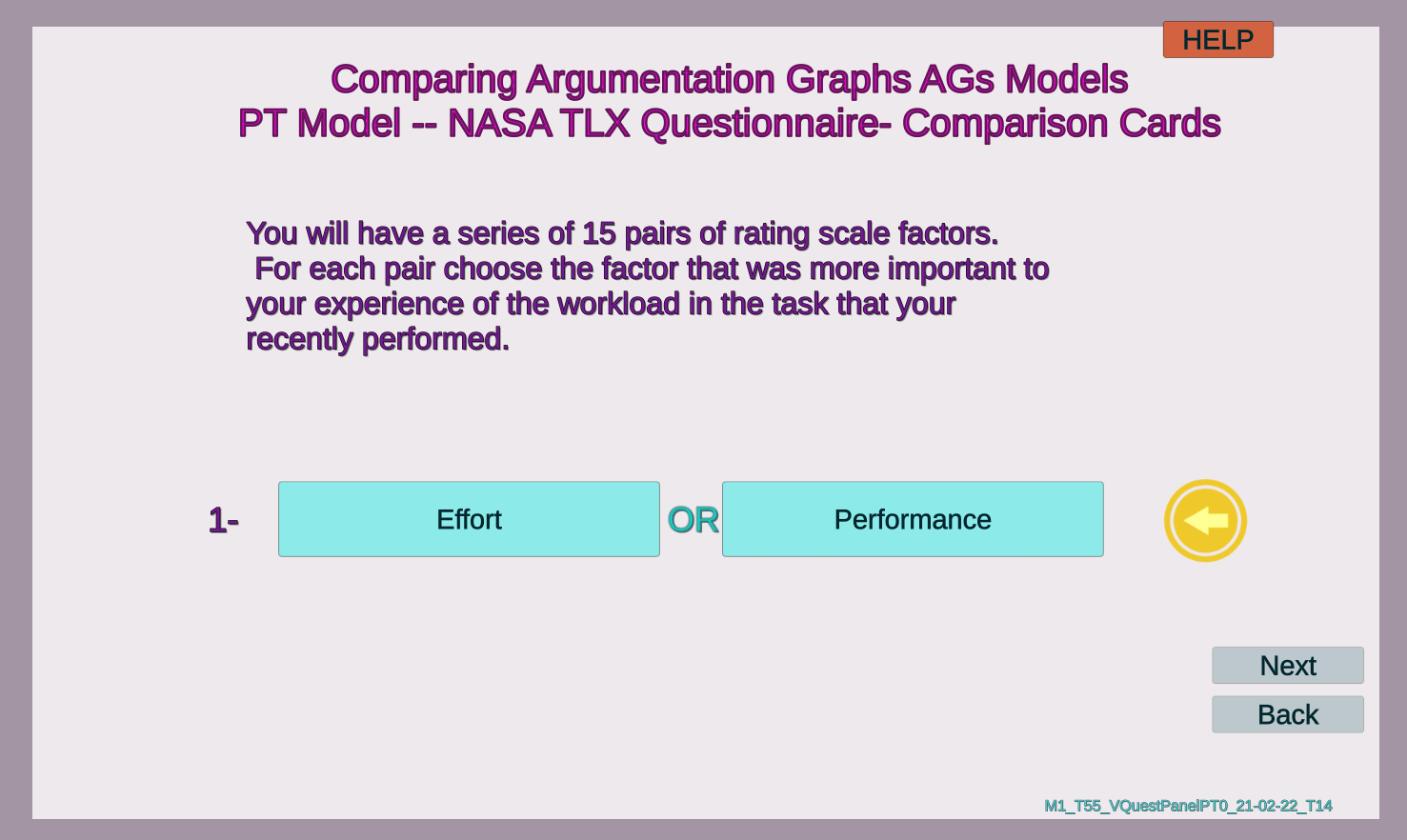} }
                \caption{NASA-TLX Questionnaire Part 1}
                \label{fig:nasa-tlx1}
             \end{subfigure}
              \begin{subfigure}{\linewidth}
                \centering
               \fbox{ \includegraphics[width=0.9\textwidth]{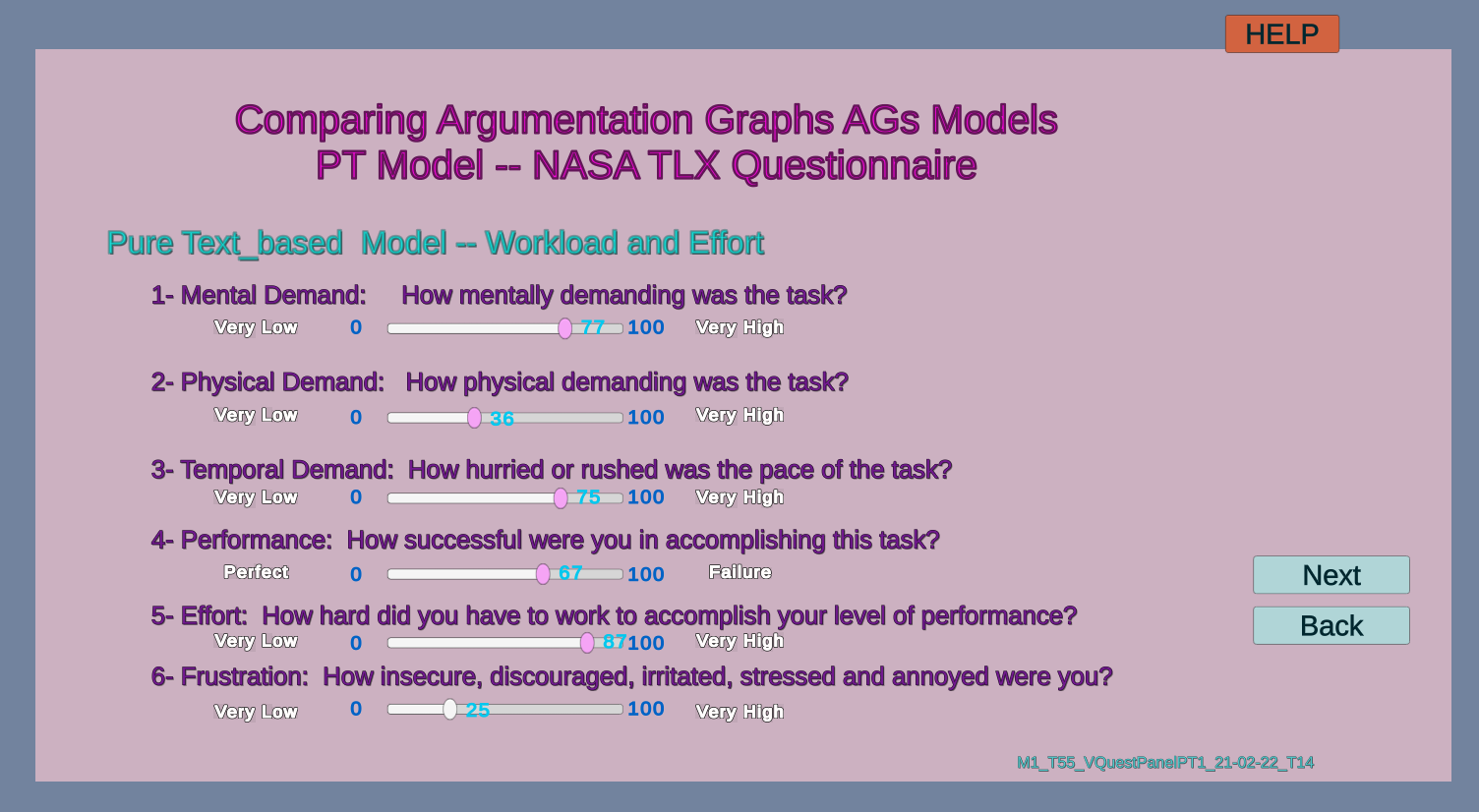}}
                \caption{NASA-TLX Questionnaire Part 2}
                \label{fig:nasa-tlx2}
             \end{subfigure}
          \caption{The second part of the NASA-TLX Questionnaire has two parts:(a) looks at each paired combination of the six scale factors to decide which is more related to their definition of workload in the task. This results in a user considering 15 paired comparisons. (b) rating each of the six factors on a scale from very low(0) to very high(100) in increments of 5. See Appendix \ref{nasa tlx sec} and section \ref{nasa_tlx_analysis}.}
       
          \label{fig:nasa-tlx}
          \end{figure*}
          
    \item \textbf{Figures \ref{fig:UEQ} are for UEQ Questionnaire Panels}
     \begin{figure*}[!ht]
           \centering
             \begin{subfigure}{\linewidth}
               \fbox{\includegraphics[width=0.9\textwidth]{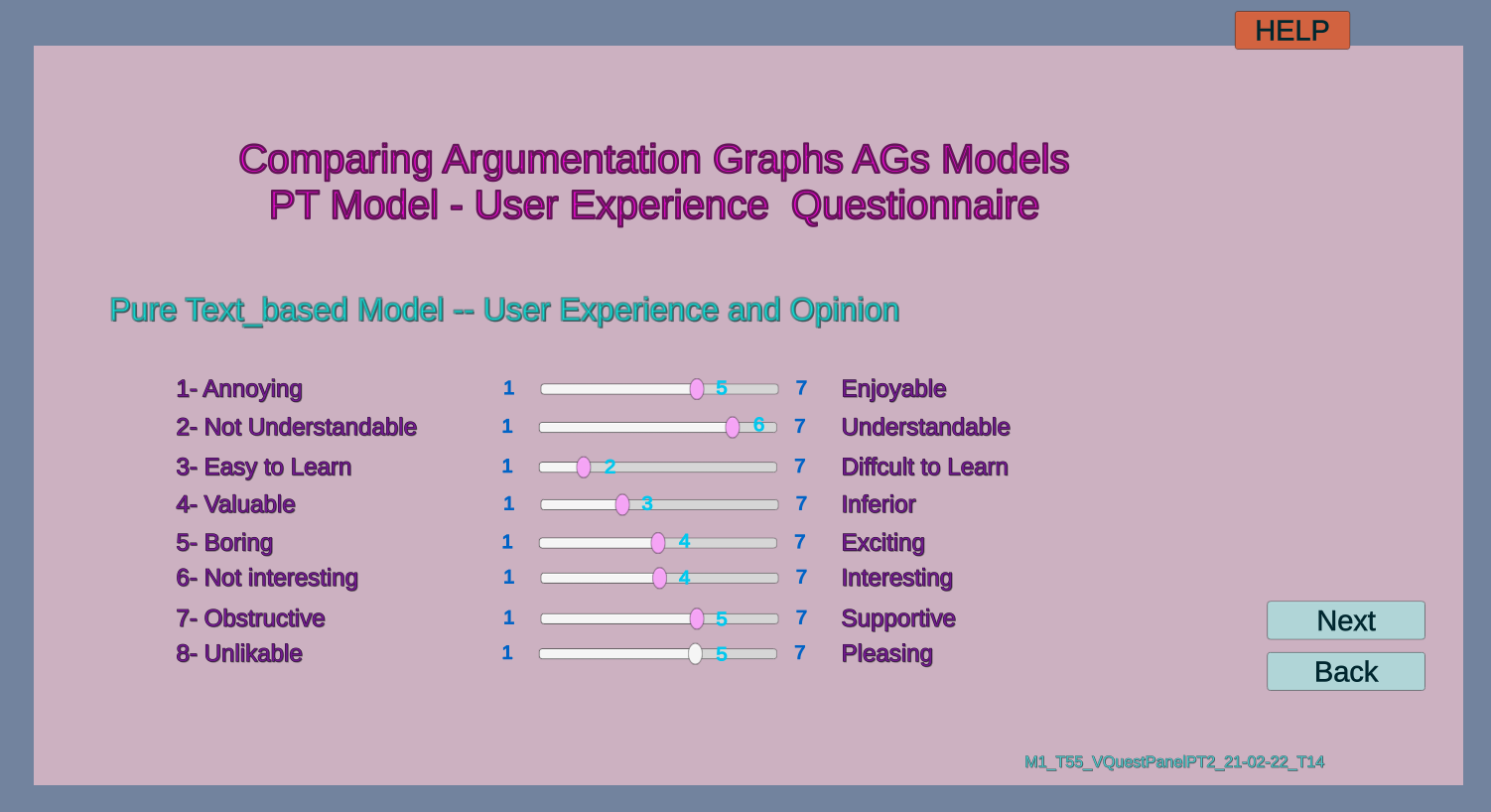}}
               \caption{UEQ Questionnaire 1-8 items }
             \end{subfigure}
              \begin{subfigure}{\linewidth}
               \fbox{ \includegraphics[width=0.9\textwidth]{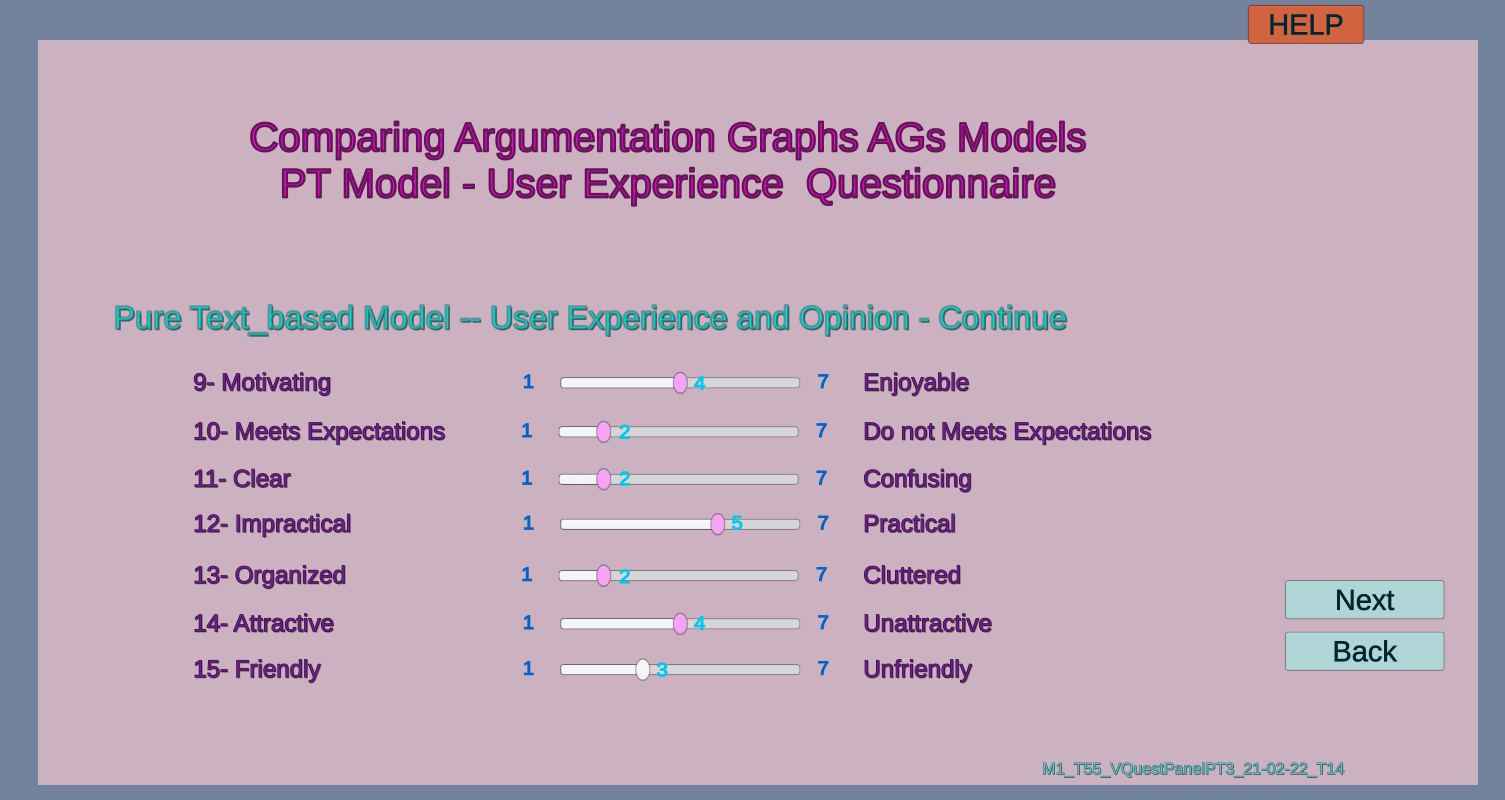}}
                \caption{UEQ Questionnaire 9-15 items }
             \end{subfigure}
          \caption{Two panels for the UEQ Questionnaire contain a total of 15 items. The items are scaled from 1 to 7, where 1 represents the most negative answer, 4 is a neutral answer, and 7 is the most positive answer. This UEQ has 15 items with 5 different main scales, which are: attractiveness, perspicuity, efficiency, dependability, and stimulation. See Appendix \ref{ueq_sec} and Section \ref{ueq_analysis}. }
        
          \label{fig:UEQ}
          \end{figure*}
          
\item \textbf{Figure \ref{fig:workflow} is for the workflow for the empirical experiment and includes references to figures, tables, and results.}
 \begin{figure*}[!ht]
           \centering
   \fbox{ \includegraphics[width=0.9\textwidth]{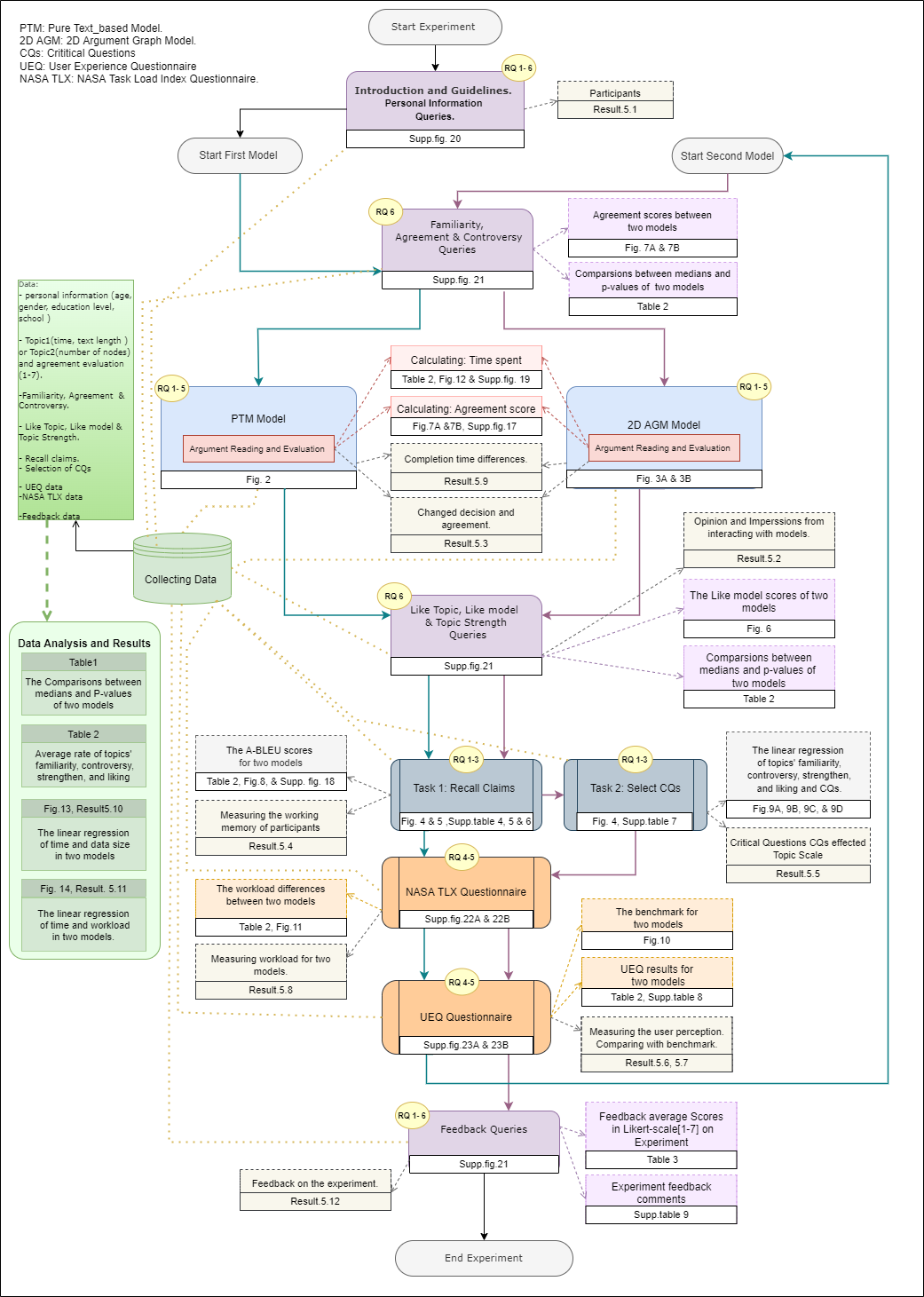}}
         \caption{The workflow for the empirical experiment includes references to figures, tables, and results.}
        
         \label{fig:workflow}
  \end{figure*}
        
\end{enumerate}

\end{document}